\documentclass[10pt,final,doublecolumn]{IEEEtran}
\usepackage{epsfig}
\usepackage{subfigure}
\usepackage{amsopn,amsmath,amssymb,amsfonts}
\usepackage{cite}
\usepackage{citesort}
\usepackage{balance}
\usepackage{float}
\newfloat{longequation}{t}{ext}

\usepackage{graphics}
\usepackage{array}
\usepackage{multirow}
\usepackage{gensymb}

\begin{document}

\title{Towards Tactile Internet in Beyond 5G Era: Recent Advances, Current Issues and Future Directions}
\author{
\authorblockN{Shree Krishna Sharma, Isaac Woungang, Alagan Anpalagan, and Symeon Chatzinotas}
\thanks{S. K. Sharma and S. Chatzintoas are with the SnT, University of Luxembourg, L-1855 Luxembourg City, Luxembourg, Emails: (shree.sharma, symeon.chatzinotas)@uni.lu.}
\thanks{I. Wounggang and A. Anpalagan are with Ryerson University, 350 Victoria Street Toronto, Canada, Emails: (iwoungan@ryerson.ca, alagan@ee.ryerson.ca).}}

\markboth{IEEE Communications Surveys \& Tutorials (Submitted Draft)} {}

\maketitle

\begin{abstract}
Tactile Internet (TI) is envisioned to create a paradigm shift from the content-oriented communications to steer/control-based communications by enabling real-time transmission of haptic information (i.e., touch, actuation, motion, vibration, surface texture) over Internet in addition to the conventional audiovisual and data traffics. This emerging TI technology, also considered as the next evolution phase of Internet of Things (IoT), is expected to create numerous opportunities for technology markets in a wide variety of applications ranging from teleoperation systems and Augmented/Virtual Reality (AR/VR) to automotive safety and eHealthcare towards addressing the complex problems of human society. However, the realization of TI over wireless media in the upcoming Fifth Generation (5G) and beyond networks creates various non-conventional communication challenges and stringent requirements in terms of ultra-low latency, ultra-high reliability, high data-rate connectivity, resource allocation, multiple access and quality-latency-rate tradeoff. To this end, this paper aims to provide a holistic view on wireless TI along with a thorough review of the existing state-of-the-art, to identify and analyze the involved technical issues, to highlight potential solutions and to propose future research directions. First, starting with the vision of TI and recent advances and a review of related survey/overview articles, we present a generalized framework for wireless TI in the Beyond 5G Era including a TI architecture, the main technical requirements, the key application areas and potential enabling technologies. Subsequently, we provide a comprehensive review of the existing TI works by broadly categorizing them into three main paradigms; namely, haptic communications, wireless AR/VR, and autonomous, intelligent and cooperative mobility systems.  Next, potential enabling technologies across physical/Medium Access Control (MAC) and network layers are identified and discussed in detail. Also, security and privacy issues of TI applications are discussed along with some promising enablers. Finally, we present some open research challenges and recommend promising future research directions.
\end{abstract}


\vspace{-10 pt}

\section{Introduction}
\label{sec:_sec1}
Towards supporting various emerging applications/platforms including Internet of Things (IoT), Virtual Reality (VR), Augmented Reality (AR), Machine-to-Machine (M2M) communications, autonomous vehicles and Tactile Internet (TI), the upcoming Fifth Generation (5G) and beyond systems are expected to achieve several performance objectives including $20$ Gbps peak data rate, $100$ Mbps data rates at the cell edges, $10^6$ devices per square kilometers, $10$ Mbps per square kilometers areal capacity and $1$ ms round-trip latency \cite{ITURvision2015}. While comparing these requirements with the current 4G systems, 5G systems need to provide $10 \times$ improvement in throughout, a $10 \times$ decrease in latency, a $100 \times$ improvement in the traffic capacity, and a $100 \times$ improvement in the network efficiency \cite{QualcomAR2018}. Out of these expected outcomes, two most challenging objectives of the upcoming 5G and beyond systems are \cite{Andrews20145G}: (i) achieving ultra-high latency (about or less than $1$ ms) and (ii) ultra-high reliability (``five-nines'' reliability, i.e., one-in-one-million chance of failure). Since these two requirements are very essential in most of the TI applications, supporting TI scenarios in wireless systems will bring unique research challenges, and thus is the focus of this paper.

The emerging TI envisions the paradigm shift of the content-oriented communications to the steer/control-based communications towards empowering people to control both the real and virtual objects via wireless channels \cite{Fettweis2014tactile,Antonakoglouhaptic}. In other words, TI is envisioned to remotely provide the real-time control and physical tactile experiences, and transform the content-delivery networks into skill-set/labor delivery networks \cite{Aijaz2016towards}. Furthermore, it is considered to be the next evolution of IoT incorporating Machine-to-Machine (M2M) and Human-to-Machine (H2M) interactions. It has a wide range of application scenarios in the industrial, eHealthCare, education and entertainment sectors, and it aims to revolutionize various aspects of our everyday life \cite{Antonakoglouhaptic}. As an example in the eHealthCare sector, a remote surgery operation can be considered, which requires the surgeon to perform the operation of a patient at the distance by feeling the sensation of the touch and providing force-feedback via haptic clothing/equipment and by implementing the required control actions and adjustments \cite{Holland2016ICT}. Another important example for the TI is a road safety scenario \cite{Fettweis2014tactile}, in which providing ultra-low latency and ultra-high reliable communications is a crucial requirement for the safety of passengers.
\subsection{Vision for TI and Recent Advances}
\label{sec:_sec11}
The TI is envisioned to be the next generation of IoT and to revolutionize various societal, economical and cultural aspects of our everyday life by enabling H2M interactions and real-time communication in Beyond 5G (B5G) networks. The TI will add various new features to the future IoT and B5G networks such as high availability, ultra-fast reaction times, high security,  carrier-grade reliability and remote control of haptic/tactile machines \cite{ITUTtechnology}. In contrast to the conventional Internet and wireless networks which usually act as a medium for audio and visual information, TI will provide a medium for the transmission of touch and actuation in the real time along with the audiovisual information. Although there are differences among the concepts of 5G, IoT networks and TI being discussed in the literature in terms of enabling devices and underlying communication paradigms, there exists a common set of requirements for future B5G networks including ultra-high reliability of about $99.999$\%, very low latency of about 1 ms, the coexistence of Human-Type-Communications (HTC) and Machine-Type-Communications (MTC) and security \cite{Maier2016TI}. The main focus of TI is on H2M communications and remote operation by utilizing various tactile/haptic devices. In addition, the emerging M2M communications is also important in TI use-cases including vehicles' electronic stability control and industrial robots \cite{ITUTtechnology}.

Furthermore, TI is expected to significantly enhance the wellbeing of human life by contributing to various sectors including personal safety, education, road traffic, energy and healthcare. Moreover, TI will contribute to address the complex challenges faced by today's human society such as increasing demand for mobility, demographic changes with the growing ageing population and transition to renewable energy production \cite{ITUTtechnology}. More specifically, TI will enable a wide range of applications and business use-cases including teleoperation systems, wireless Augmented Reality (AR)/Virtual Reality (VR), remote driving, cooperative automotive driving, smart energy systems, Internet of drones, eHealthcare, intelligent traffic control, and tactile robots. In addition, various mission critical applications including transportation, healthcare, mobility, manufacturing and transportation, and also the non-critical applications including edutainment and events can be considered promising use-cases of TI \cite{Holland2019IEEE}.

However, existing network infrastructures are conceptually and technically insufficient to support the emerging TI applications.  TI brings various non-conventional demands and stringent requirements in future B5G networks in terms of reliability, latency, sensors/actuators, access networks, system architecture and mobile edge-clouds \cite{ITUTtechnology}. The design requirements of TI systems/devices to achieve real-time interactions are dependent on the participating human senses since the speed of human's interaction with the underlying system/environment is limited by our perceptual processes. Also, it is crucial to adapt the feedback of the system based on human reaction time.

In addition to very low end-to-end latency and ultra-high reliability design requirements of the TI, data security, availability and dependability of systems need to be ensured without violating the low-latency requirements while also taking into account of encryption delays. Nevertheless, existing centralized architectures are not sufficient to meet these requirements  and more distributed network architectures based on mobile-edge computing and cloudlets need to be investigated to bring the TI applications closer to the end-users \cite{Maier2016TI}. Also, it is essential to  redesign future wireless access networks by investigating novel resource allocation, feedback mechanisms, interference management and medium-access control techniques in order to meet the stringent reliability and latency requirements of TI applications \cite{ITUTtechnology}. In addition to the enhancement in various aspects of physical and Medium Access Control (MAC) layers, emerging network technologies including Software Defined Networking (SDN), Network Function Virtualization (NFV), network coding and edge/fog computing are considered promising paradigms for supporting the TI applications in future B5G networks \cite{Bojkovic2017vision}.

Over the last decade, there has been a significant research attention in developing smart tactile sensing systems, which are mainly composed of tactile sensors (capacitive, piezo-resistive, piezoelectric and optical) and intelligent signal processing tools having the capability of information interpretation and taking decisions \cite{Zou2017novel}. However, several interdisciplinary research efforts are needed in designing smart tactile sensing systems in terms of device materials, fabrication technologies, signal processing and Machine Learning (ML) algorithms to deal with the complex multi-dimensional data from the tactile sensors. In addition, the transmission of tactile sense/data over the wireless Internet while enabling the remote touch feeling as well as the remote control, is another crucial aspect to be addressed. In addition, various aspects of underlying wireless communication networks such as reliability, latency, security and availability need to be taken into the design of TI systems. Several standardization works are expected in the near future to create synergies between the TI's application areas and the evolution of existing wireless networks.   From the regulatory standards perspective,  by considering TI  under a ``Technology Watch'' area, the telecommunication standardization sector of International Telecommunication Union (ITU) has published a report on various aspects of TI including the applications in both the mission-critical and non-critical areas, the advantages for the society and  implications for the equipment \cite{Fettweis2014}.

In terms of the ongoing standardization, a new standard family IEEE P1918.X ``Tactile Internet'' has been defined \cite{Dresslertut2017}. The scope of the baseline TI standard IEEE P1981.1 is to define a TI framework, incorporating the descriptions of its definitions and terminology, including the necessary functions and technical assumptions, as well as the application scenarios \cite{Holland2019IEEE}. Within this, IEEE P1918.X defines the architecture technology and assumptions in TI systems, IEEE P1918.X.1 is dedicated for Codecs for the TI, IEEE P1918.X.2 is focussed on Artificial Intelligence (AI) for Tactile Internet and IEEE P1918.X.3 is working towards MAC for the TI. Also, the Industrial Internet Consortium (IIC) (initiated by GE, CISCO, AT\&T, Intel and IBM, and which later supported by 160 companies) is developing a standard for low-latency Industrial IoT (IIoT) for different smart cyber-physical systems including smart transportation systems, smart manufacturing,  and smart healthcare systems \cite{Szymanski2016TI,WEIIT15}.

\begin{table*}
\caption{\small{Classification of survey/overview works in the areas of IoT/mMTC and Tactile Internet including its applications such as haptic communications and wireless VR/AR.}}
	\centering
\begin{tabular}{|l|l|l|}
\hline
Main domain & Sub-topics & References \\ \hline
  & Enabling technologies/protocols, applications and challenges & \cite{Fuqaha2015IoT,Wang2017survey,Ghavimi2015M2M,Bockelmann2016,Wang2016cellular,Dawy2017towards,Alvarino2016overview,Elsaadany2017cellularLTEA,Hoglund2017overview} \\
& Random access schemes  &  \cite{Layarandomaccess2014,Yangnarrwoband2017,Ali2017LTE} \\
IoT/mMTC/M2M/URLLC  & Traffic characterization and issues   & \cite{Soltanmohammadi2016} \\
& Transmission scheduling     & \cite{Gotsis2012M2M,Mehaseb2016classification} \\
& IoT big data analytics   & \cite{Marjani2017bigdata,Verma2017survey,SKSIEEE2017} \\
& Short packet transmission & \cite{Durisi2016IEEEproc} \\
& Latency reduction techniques & \cite{Parvez2018survey,Briscoe2016internetlatency,Shariatmadari20185G} \\
\hline
& Vision, applications and challenges  & \cite{Fettweis2014tactile,Maier2016TI} \\
Tactile Internet & 5G-enabled TI & \cite{Simsel20165G} \\
& Haptic communications & \cite{Steinbach2012haptic,Antonakoglouhaptic,Challengeshapticcommunication,Steinbach2011haptic} \\
& Wireless virtuality/augmented reality & \cite{Bastug2017commun} \\
\hline
\end{tabular}
	\vspace{-15 pt}
\label{tab: referencesclassification}
\end{table*}

\subsection{Review of Related Overview/Survey Articles}
\label{sec:_sec12}
There exist several overview and survey papers in the areas of IoT, massive Machine Type Communications (mMTC), M2M communications, Ultra Reliable and Low Latency Communications (URLLC), but only a few in the area of TI. In Table \ref{tab: referencesclassification}, we list the related topics under these domains and the existing overview/survey articles. The existing survey and overview articles \cite{Fuqaha2015IoT,Wang2017survey,Ghavimi2015M2M,Bockelmann2016,Wang2016cellular,Dawy2017towards,Alvarino2016overview,Elsaadany2017cellularLTEA,Hoglund2017overview}
have discussed various aspects of IoT/mMTC/M2M and URLLC systems including enabling technologies/protocols, applications and challenges. Furthermore, the article \cite{Gotsis2012M2M} discussed various challenges and future perspectives for scheduling M2M transmissions over LTE networks and another article \cite{Mehaseb2016classification} provided a comprehensive classification of transmission scheduling techniques from the M2M communications perspective. Moreover, some survey and overview articles  provided a comprehensive discussion on various aspects of IoT data analytics including the related architectures \cite{Marjani2017bigdata}, network methodologies for real-time data analytics \cite{Verma2017survey} and collaborative edge-cloud processing \cite{SKSIEEE2017}. In addition, the authors in  \cite{Durisi2016IEEEproc} provided an overview of recent advances in information theoretic principles towards characterizing the transmission of short packet transmissions in IoT/mMTC systems.

In the direction of latency reduction techniques, which is very crucial for TI applications, a few overview and survey papers exist \cite{Parvez2018survey,Briscoe2016internetlatency,Shariatmadari20185G}. The authors in \cite{Parvez2018survey} provided a detailed survey on the emerging technologies to achieve low latency communications by considering different domains including Radio Access Network (RAN), core network and caching network. Another article \cite{Briscoe2016internetlatency} provided a thorough classification of latency reduction techniques based on the sources of delay along with a comparison of the merits of different techniques. In addition, the article \cite{Shariatmadari20185G} reviewed the URLLC requirements for 3GPP LTE Releases 14 and 15 and 5G New Radio (NR), and discussed novel enhancements to design the control channels towards supporting URLLC.

In the area of TI, the authors in \cite{Fettweis2014tactile} provided a brief overview of emerging TI application scenarios mainly in the areas of healthcare, mobility, education, smart grids and manufacturing. Another overview article in \cite{Maier2016TI} discussed the vision, recent advances and open challenges of TI by focusing on the reliability and latency performance gains of Fiber-Wireless (FiWi) enhanced LTE-Advanced networks. Furthermore, the article \cite{Simsel20165G} presented the technological concepts underlying the intersection of emerging 5G systems and TI scenarios, and outlined the key technical requirements and architectural methods for TI regarding wireless access protocols, radio resource management and edge-cloud capabilities.

In the context of haptic communications, there exist a few overview and survey papers in the literature  \cite{Steinbach2012haptic,Antonakoglouhaptic,Challengeshapticcommunication,Steinbach2011haptic}. The authors in \cite{Steinbach2012haptic} provided a  state-of-the-art review in haptic communications both from the technical and psychophysical perspectives, and also discussed the need of various objective quality metrics for haptic communications. Another article \cite{Antonakoglouhaptic} provided a survey of the methodologies and technologies for incorporating haptic communications in the 5G TI systems, and also discussed various aspects related to haptic tele-operation systems, haptic data reduction and compression, haptic control systems and haptic data communication protocols. Furthermore, the overview article \cite{Steinbach2011haptic} discussed the state-of-the-art and challenges of haptic data compression and communication for teleaction and telepresence operations. In addition, the authors in \cite{Challengeshapticcommunication} provided a summary of the requirements and challenges of haptic communications along with some possible solutions.

With regard to wireless AR/VR, the authors in \cite{Bastug2017commun} highlighted the importance of VR as a disruptive use-case for 5G and beyond wireless systems, and subsequently discussed potential research avenues and scientific challenges for the implementation of interconnected VR over wireless links. Also, three different VR case studies related to joint resource allocation and computing, proactive versus reactive computing, and AR-enabled self-driving vehicles, were presented.

\begin{table*}
\caption{\small{Definitions of Acronyms}}
\centering
\renewcommand{\arraystretch}{0.9}
\begin{tabular}{ll|ll}
  \hline 
 \textbf{Acronyms}  & \textbf{Definitions} & \textbf{Acronyms}  & \textbf{Definitions} \\
 \hline
AR & Augmented Reality & M2M & Machine-to-Machine \\
 AI & Artificial Intelligence & MTC & Machine-Type-Communications \\
AS & Application Server  & MAC & Medium Access Control \\
ACK  & Acknowledgement& mMTC & massive Machine-Type-Communications \\
 ARQ  & Automatic Repeat Request & MCS  & Modulation and Coding Scheme  \\
BLER & Block-Error-Rate  & NACK & Non-Acknowledgement\\
BS & Base Station & NFV & Network Function Virtualization \\
B5G & Beyond Fifth Generation & NR  & New Radio \\
CSI & Channel State Information  & NCS & Networked Control Systems \\
C-Tx & Concurrent Transmission & OOB & Out-of-Band  \\
CU  & Central Unit &  OFDM  & Orthogonal Frequency Division Multiplexing \\
CQI  & Channel Quality Indicator  & POL  & Passive Optical Local Area \\
D2D & Device-to-Device & QoT & Quality-of-Task \\
DSSS & Direct-Sequence Spread Spectrum &  QoE & Quality-of-Experience\\
eMBB & Enhanced Mobile Broadband & QoS  & Quality of Service \\
FiWi  & Fiber-Wireless & RU  & Radio Unit \\
FBMC  & Filter Bank Multi-Carrier & RL & Reinforcement Learning \\
HD   & High Definition & RB & Resource Block \\
HSI  & Human System Interface &  RAT  & Radio Access Technology  \\
H2H  & Human-to-Human & SDN & Software-Defined Networking \\
H2M  & Human-to-Machine & SCMA & Sparse Code Multiple Access \\
HTC  & Human-Type-Communications   & SNR & Signal-to-Noise Ratio \\
HMD  & Head Mounted Device & SR & Scheduling Request \\
HARQ & Hybrid ARQ  & TI  & Tactile Internet \\
IoT  & Internet of Things & TTI & Transmission Time Interval \\
IoE  & Internet of Energy & TCP & Transport Control Protocol \\
IVR  & Immersive Virtual Reality  & TWDM & Time and Wavelength Division Multiplexing \\
ICN  & Information-Centric Networking & UAV & Unmanned Aerial Vehicle \\
IIoT & Industrial IoT  & UH & Ultra-High \\
LTE & Long-Term Evolution & UFMC  & Universal Filterbank Multi-Carrier  \\
LTE-A & Long-Term Evolution-Advanced  & URLLC & Ultra-Reliable and Low-Latency Communications \\
LAN  & Local Area Network  &  V2I & Vehicle-to-Infrastructure   \\
MEC & Mobile Edge Computing  & V2V & Vehicle-to-Vehicle   \\
MTP & Motion-to-Photon & VR & Virtual Reality \\
MIMO & Multiple-Input Multiple-Output & V2X  & Vehicle-to-Anything \\
ML & Machine Learning & WBAN & Wireless Body Area Network \\
\hline 
\end{tabular}
\label{tab: Acronyms}
\vspace{-10 pt}
\end{table*}

\begin{figure*}
	\begin{center}
		\includegraphics[width=4.5 in]{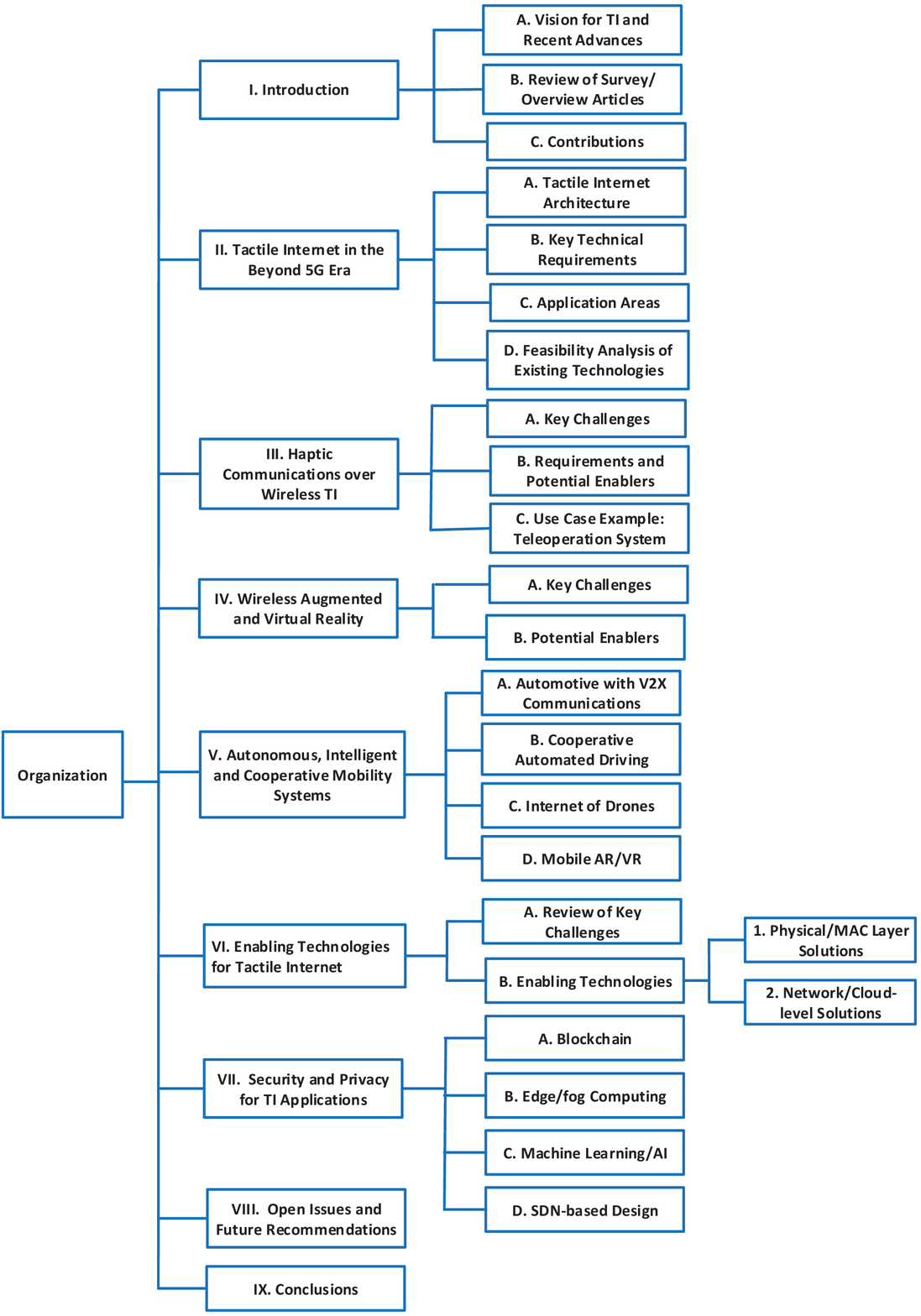}
		\caption{\small{Structure of the paper.}}
		\label{fig: paperstructure}
	\end{center}
	\vspace{-20 pt}
\end{figure*}

\subsection{Contributions of the Paper}
\label{sec:_sec13}
Although there exist a few surveys and overview articles in the areas of TI, they fail to provide a holistic view on wireless TI along with the issues involved with the transmission of tactile traffic in addition to the conventional audio-visual and data traffics over the complex wireless communications environment. In this direction, the main objectives of this survey paper are: to provide a comprehensive view on wireless TI along with a generalized TI framework (Fig. \ref{fig: TIframwork} in Section \ref{sec:_sec2}), to discuss current issues towards supporting TI applications in 5G and beyond wireless networks, to provide a detailed review of recent advances in the 5G and beyond TI use-cases, to propose potential enabling technologies for wireless TI, to identify open issues and to suggest future research directions. In the following, we briefly highlight the main contributions of this survey paper.
\begin{enumerate}
\item We propose a generalized framework for wireless TI, identify its main components, and subsequently discuss various aspects of wireless TI including a basic TI architecture, key technical requirements, main application areas and the feasibility analysis of existing technologies.
\item By categorizing the B5G TI use-cases into three domains, namely, haptic communications, wireless AR/VR, and autonomous, intelligent and cooperative mobility systems, we provide their comprehensive state-of-the-art review along with the main challenges, key enablers and some example use-cases.
\item We identify the key challenges to be addressed while supporting TI applications over wireless media, and subsequently propose and discuss various potential enabling technologies across the physical/MAC and network layers including network edge and cloud-level techniques in detail.
\item Security and privacy threats to be faced by TI applications are discussed along with the related state-of-the-art, and also some promising techniques/architectures for security enhancements in TI systems are suggested.
\item Several open research issues are identified and some interesting research directions are provided to stimulate future research activities in the related domains.
\end{enumerate}

\subsection{Paper Organization}
\label{sec:_sec14}
The remainder of this paper is structured as follows: Section \ref{sec:_sec2} presents a generalized framework for TI in the beyond 5G era and discusses various aspects of TI including TI architecture, key technical requirements, application areas and the feasibility analysis of existing technologies. Section \ref{sec:_sec3} focuses on various aspects of haptic communications including underlying challenges, requirements, potential enablers and a use-case example on teleoperation system along with a thorough review of the existing literature. Section \ref{sec:_sec4} provides the details on the wireless augmented and virtual reality along with the key technical challenges and the main enablers while Section \ref{sec:_sec5} includes a thorough review of various autonomous, intelligent and cooperative mobility systems, mainly, automotive with V2X communications, cooperative automated driving, Internet of drones and mobile AR/VR. Section \ref{sec: sec6} discusses research challenges and various enabling technologies for TI including physical, MAC and network layer solutions, and cloud-level techniques while Section \ref{sec:_sec7} discusses security and privacy aspects of TI applications. Section \ref{sec:_sec8} discusses open issues and future recommendations on various related topics. Finally, Section \ref{sec:_sec9} concludes the paper.  To improve the flow of this manuscript, we have provided the definitions of acronyms in Table \ref{tab: Acronyms} and the structure of the paper in Fig. \ref{fig: paperstructure}.

\section{Tactile Internet in the Beyond 5G Era}
\label{sec:_sec2}
\begin{figure*}
	\begin{center}
		\includegraphics[width=6.5 in]{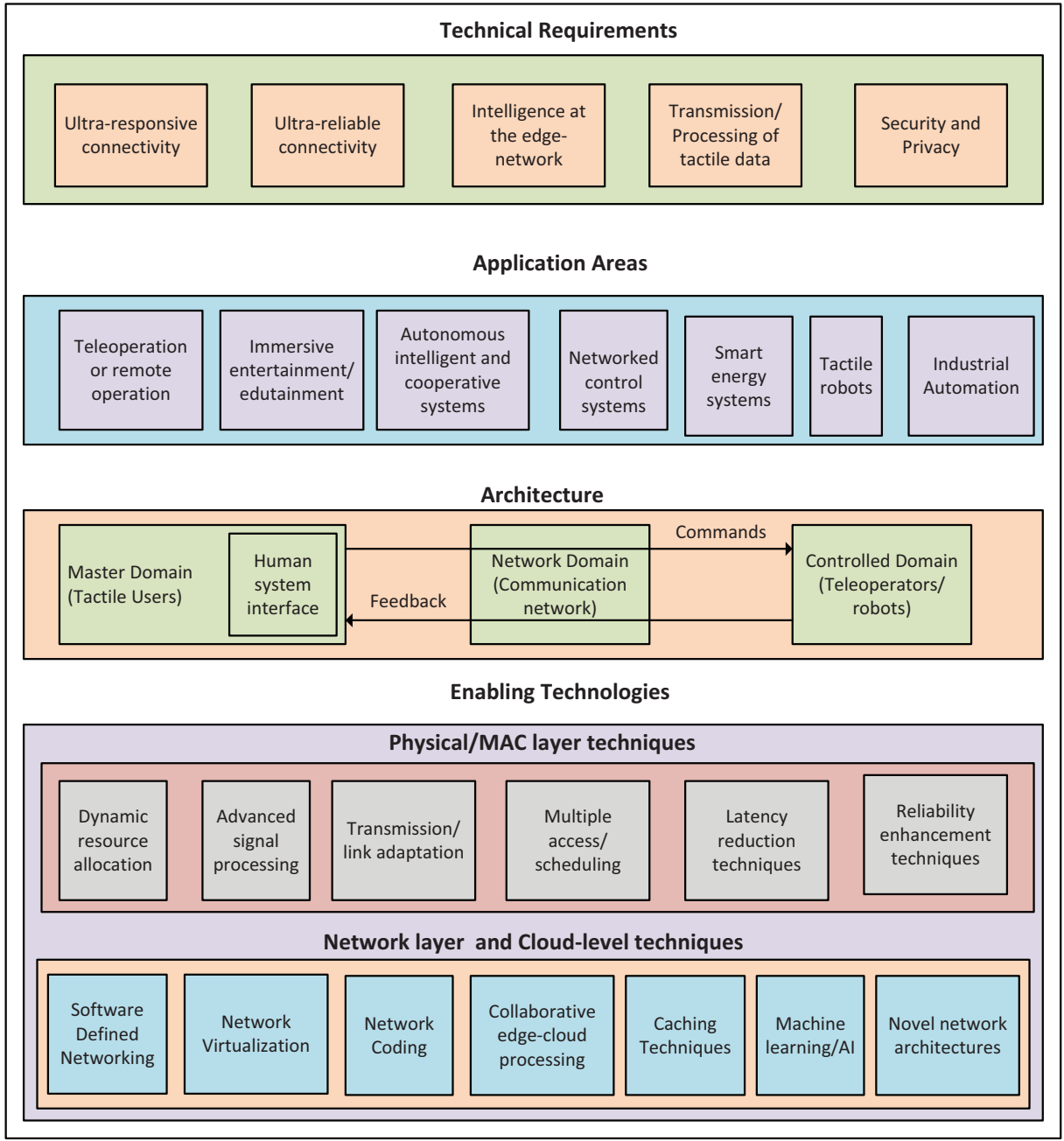}
		\caption{\footnotesize{A generalized framework for Tactile Internet in the Beyond 5G Era.}}
	\label{fig: TIframwork}
	\end{center}
\end{figure*}

In Fig. \ref{fig: TIframwork}, we  present a generalized framework comprising various aspects of TI including key technical requirements, main application domains, a basic architecture and enabling technologies. The key technical requirements of TI include ultra-responsive connectivity, ultra-reliable connectivity, intelligence at the edge network, efficient transmission and low-complexity processing of tactile data, which are elaborated later in Section \ref{sec:_sec22}. Regarding the applications of TI, the main areas include tele-operation or remote operation, immersive entertainment/edutainment, autonomous intelligent and cooperative systems, networked control systems, smart energy systems, tactile robots and industrial automation, which are discussed in Section \ref{sec:_sec23}. As depicted in Fig. \ref{fig: TIframwork}, the basic architecture of TI is composed of a master domain, a network domain and a controlled domain, which are described later in Section \ref{sec:_sec24}. Furthermore, several physical/MAC layer and cloud/network level technologies highlighted in Fig. \ref{fig: TIframwork} are detailed in Section \ref{sec: sec6}.

In the following subsections, we present a generalized architecture of TI,  the key technical requirements, the main applications areas of TI, and a brief discussion on the state-of-the-art work related to the analysis of existing technologies with regard to the technical requirements of TI.

\subsection{Tactile Internet Architecture}
\label{sec:_sec24}
In contrast to the conventional Internet utilized for transmitting audio and video information, TI envisions to transmit touch and actuation information in addition to the audiovisual information. From the communication perspective, the main difference is that the feedback in an TI system can be in terms of various kinesthetic and vibro-tactile parameters such as position, motion, vibration and surface texture in addition to audio/vidual feedback, thus forming a global control loop while the conventional system utilizes only the audio/visual feedback without any control loop \cite{Simsel20165G}. Also, the haptic sense in TI applications occurs bilaterally, for example, in a telepresence system, the motion (i.e., velocity or position) is transmitted to the tele-operator and the force/torque from the environment is sent back to the Human System Interface (HSI) \cite{Steinbach2012haptic}.

\begin{figure*}
	\begin{center}
		\includegraphics[width=6.5 in]{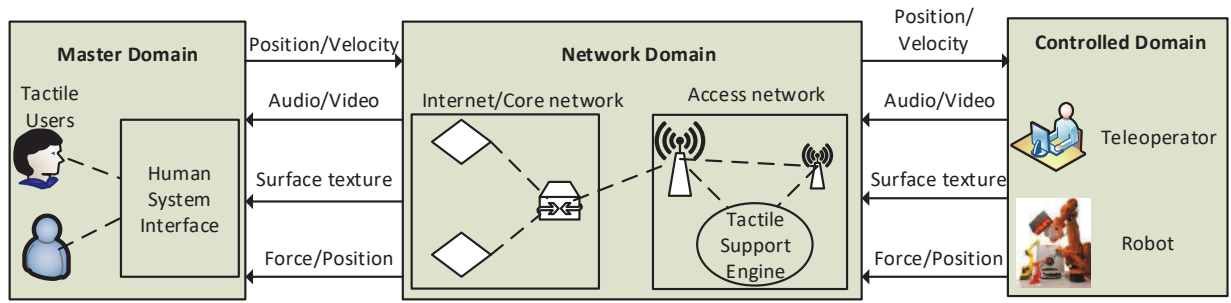}
		\caption{\footnotesize{A generalized architecture of a wireless Tactile Internet system.}}
	\label{fig: TIarchi}
	\end{center}
\end{figure*}

In Fig. \ref{fig: TIarchi}, we present a generalized architecture of a TI system, which mainly comprises a master domain, a network domain and a controlled domain. The master domain generally consists of a human operator (tactile user) and the HSI, which is responsible for converting the human input to the tactile data by utilizing suitable tactile encoding techniques. The tactile data generated by the HSI is transmitted to the controlled domain via the network domain.  The controlled domain or environment comprises a remotely controlled robot or teleoperator and with the help of various command signals (i.e., velocity, position), the master domain directly controls the controlled domain. The controlled domain then provides feedback signals (i.e., force/position, surface texture) to the master domain. In addition to the haptic feedback signals, the master domain also receives audio/visual feedback signals from the controlled domain. The master domain and controlled domains are connected  via a two-way communication link over the network domain with the help of various command and feedback signals, forming a global control loop. Out of the aforementioned components of TI, the main focus of this paper will be on the network domain, i.e., communication network.

The  network domain of the TI system acts as the communication channel between tactile users (master domain) and the remotely controlled environment (controlled domain). This communication domain may be composed of an Internet/core network, a Radio Access Network (RAN), and a tactile support engine as depicted in Fig. \ref{fig: TIarchi}. Furthermore, the underlying communication channel needs to satisfy various requirements such as ultra-responsive connectivity and ultra-reliable connectivity highlighted later in Section \ref{sec:_sec22}. The upcoming B5G networks should be adapted to satisfy these requirements towards supporting emerging TI applications. The main functions of B5G core network to support a TI system include handling of edge-cloud interactions and access, application-aware Quality of Service (QoS) provisioning and security. Similarly, the key functions of B5G RAN in the context of TI system include the efficient support of heterogeneous Radio Access Technologies (RATs) including the conventional LTE/LTE-A based cellular, 5G New Radio and emerging B5G technologies,  radio resource management and QoS aware scheduling of TI users, the coexistence of TI applications with other verticals including Vehicle-to-Vehicle (V2V), Vehicle-to-Infrastructure (V2I), smart grid, M2M, and provisioning of reliable packet delivery. Furthermore, the transmission and synchronization of heterogeneous data streams via the communication medium may face challenges due to packet loss, higher packet rate and variable delay, demanding for intelligent packet-switched network protocols capable of analyzing the network situations, synchronizing the data streams and provisioning the required QoS \cite{Antonakoglou2018haptic}.

\subsection{Key Technical Requirements}
\label{sec:_sec22}
The main technical requirements for the TI include the following \cite{Simsel20165G,Li20195Gbased}.
\begin{enumerate}
\item \textbf{Ultra-responsive Connectivity}: Most of the TI applications require the end-to-end latency/round trip delay to be in the order of about $1$ ms. The end-to-end latency refers to the summation of the transmission times required while sending the information from a sensor/device (or human for the case of haptic communication) via the communication infrastructure to a control server, the information processing time at the server, the processing at different communication hops (i.e., routers, switches), and the retransmission times via the communication infrastructure back to the end-device (or human).
\item \textbf{Ultra-reliable Connectivity}: Another important requirement for the TI is the ultra-reliable network connectivity, in which reliability refers to the probability to guarantee a required performance under given system constraints and conditions over a certain time interval. As an example, factory automation scenario in a smart factory demands a reliability of about $99.999$\% for about $1$ ms latency \cite{Yilmaz2015analysis}. One of the potential solutions to enhance the reliability for TI applications is to employ concurrent connections with multiple links \cite{Ohmann2014highavailability}, and also to utilize multiple paths for graph connectivity to avoid a single point of failure. However, this approach depends on the channel dynamics and the availability of Channel State Information (CSI) knowledge. Including higher Signal-to-Noise Ratio (SNR) margins in the link budget and employing stronger channel coding are also important solutions towards enhancing the reliability of a communication link. Improving the reliability will also help to reduce the latency due to the lower number of resulting retransmissions.
\item \textbf{Distributed Edge Intelligence}: Suitable Artificial Intelligence (AI)/ML techniques need to be investigated to be employed at the edge-side of the wireless TI networks in order to facilitate the interpolation/extrapolation of human activities and predictive caching for reducing the end-to-end latency. Furthermore, AI/ML-based predictive actuation methods need to be investigated in order to enhance the range of tactile services/applications.
\item \textbf{Transmission and Processing of Tactile Data}: To facilitate the transmission of tactile information over the packet-switched networks, tactile encoding mechanisms need to be developed. Also, to handle the highly multi-dimensional nature of human tactile perception, an effective audio/visual sensory feedback mechanism needs to be investigated.
\item \textbf{Security and Privacy}: Other key requirements of the TI are security and privacy under strict latency constraints. To fulfill these requirements, physical layer security techniques with low computational overhead, secure coding techniques, and reliable and low-latency methods to identify the legitimate receivers, need to be investigated.
\end{enumerate}

\subsection{Application Areas}
\label{sec:_sec23}
As highlighted earlier, the TI is expected to influence various aspects of human society and has a wide range of application areas ranging from eHealthCare to industrial automation. Some of the application scenarios specified in the literature include \cite{Antonakoglouhaptic,Antonakoglouhaptic}: (i) Haptic communications, (ii) Augmented Reality (AR)/Virtual Reality (VR), (iii) Remote monitoring and surgery, (iv) Wireless controlled exoskeletons, (v) Remote education (tele-mentoring), (vi) Remote driving/autonomous vehicles, (vii) Traffic control, (viii) Industrial tele-operation including robotics and manufacturing, (ix) Smart grid, and (x) Smart city.

Furthermore, the baseline TI standard IEEE 1918.1 has provided the key features and performance metrics of the following use-cases \cite{Holland2019}: (i) Immersive Virtual Reality (IVR), (ii) teleoperation, (iii) automotive, (iv) Internet of drones, (v) interpersonal communications, (vi) live haptic-enabled broadcast and (vii) cooperative automotive driving. For example, for IVR applications, the traffic type for the slave (haptic users) to master (IVR systems) is haptic feedback while in the direction of master to slave, video, audio and haptic feedback traffic types are involved. For this IVR application, the latency requirement for haptic feedback in the slave to master direction is $<5$ ms while in the other direction (master to slave), video and audio have the latency requirement of $<10$ ms and for haptic feedback, the latency requirement is  $1-50$ ms \cite{Holland2019}. The aforementioned main TI applications can be categorized under the following headings \cite{AIjazproc2019TI,Simsel20165G}.

\textbf{1. Teleoperation or remote Operation Systems}: These systems can perform the manipulation of tasks in distant/inaccessible and hazardous places by enabling the interaction of humans with real/virtual objects. With the help of haptic information, various aspects such as motion, vibration, texture and force can be communicated to create the feeling of being present in the remote environment. In contrast to the conventional wired solutions over short distance, wireless teleoperation systems face challenges in terms of meeting the strict requirements of latency and reliability for timely and reliable remote interaction. To this end, TI finds an important application in enabling the remote interactions over long distances via wireless links.

\textbf{2. Immersive Entertainment and Edutainment Systems}: Another important application of TI is the entertainment industry where the TI can enable a range of new immersive entertainment services via real-time transmission of multi-sensory information including audio, video and haptic. Also, new gaming applications will benefit from the
emerging VR and AR systems. Furthermore, live haptic broadcasting over TI can provide the feeling of live event to the end-users \cite{Guit2015tactile}. Moreover, by enabling the haptic interaction between teachers and students, TI can provide a completely new method of education, thus enabling new edutainment applications. However, this leads to the challenge of having a multimodal (auditory, visual and haptic) H2M interface capable of operating with very low end-to-end latency.

\textbf{3. Autonomous, Intelligent and Cooperative Mobility Systems}: Under intelligent mobility systems, the main applications of TI include autonomous driving, remote driving, vehicle platooning, virtually coupled train systems, aerial drones and Unmanned Aerial  Vehicles (UAVs). Vehicle-to-Vehicle (V2V) and Vehicle-to-Infrastructure (V2I) communication systems are essential for self-driving capabilities in future wireless systems. The aforementioned autonomous, intelligent and cooperative mobility systems require future TI systems to meet the stringent constraints in terms of latency, reliability and stability control.

\textbf{4. Wireless Networked Control Systems (NCS)}: Wireless NCSs are distributed systems consisting of sensors, actuators, and controllers, in which the control and feedback signals are exchanged over a common communication network with the help of a global control loop with strict latency constraints. The NCSs face the challenge of instability in wireless channels due to the involved time delays and packet losses and TI may be a promising paradigm to enable wireless NCSs by providing highly reliable and ultra-low latency connectivity solutions.

\textbf{5. Smart Energy Systems}: With the recent advances in smart grid technologies, the existing energy ecosystem is being transferred to Internet of Energy (IoE), which can enable various operations such as distributed energy generation, interconnected storage network and demand-side response \cite{AIjazproc2019TI}.
The TI may provide ultra-responsive connectivity to meet the challenge of achieving high dynamic control over the distributed energy supply units since out-of-phase synchronization power injection may result in the unusable reactive power \cite{ITUTtechnology}.

\textbf{6. Tactile Robots}: With the recent advances in electronics, human-machine interaction technologies, AI techniques, and haptic devices, the recently emerged kinesthetic soft robot is expected to evolve its next generation, i.e., Tactile Robot. Tactile Robot is envisioned to provide fully immersive remote representation of a human enhanced with the AI and with the capability of performing physical interactions and high-fidelity feedback to a human operator over significant distances \cite{Haddadin2019TI}. Such a feedback from the tactile robot to the human is provided with the help of connected wearables and various modules such as tactile sensing, temperatures, pressure and audiovisual perception. Over the current generation of soft robots, tactile robot provides several advantages including safety algorithms, interoceptive and exteroceptive tactile feedback and enhanced quality metrics \cite{Haddadin2019TI}.

\textbf{7. Industrial Automation}: The industrial automation in smart factories is one of the key application areas of the TI. The emerging industrial revolution, i.e, Industry 4.0 consisting of storage systems, smart machines and production facilities, is expected to have the capabilities of real-time response, autonomous information exchange over wireless, increased flexibility, and self-organization \cite{Simsek20165Gtactile}. In contrast to the today's control systems, future industrial control systems are expected to be fully or partially operating over wireless links to enhance the production flexibility. This causes several issues to be addressed in terms of reliability, latency, energy consumption and security. Various industrial applications may have different data rate, end-to-end latency and security requirements \cite{Weiner2014lowlatency}. The emerging TI solutions may be significantly useful in meeting these requirements towards enabling full automation, and highly agile and flexible production process in smart factories.

The aforementioned applications of TI can be broadly considered under the following three main paradigms: (i) haptic communications, (ii) wireless augmented and virtual reality, and  (iii) autonomous, intelligent and cooperative mobility Systems, which are discussed in detail along with the existing state-of-the-art review in Section \ref{sec:_sec3}, Section \ref{sec:_sec4} and Section \ref{sec:_sec5}, respectively. In Table \ref{tab: TIsuecases}, we have listed the main research themes under these categories and the corresponding references.

\begin{table*}
\caption{\small{Different use-cases of tactile communications in 5G and beyond wireless networks and corresponding references.}}
\centering
\begin{tabular}{|l|l|l|}
\hline
\textbf{TI Use Cases} & \textbf{Main Theme} &  \textbf{References}  \\
\hline
 & Vision, challenges and applications & \cite{Steinbach2012haptic,Challengeshapticcommunication,Aijaz2017tactile} \\
Haptic Communications & Radio resource allocation  &  \cite{Aijaz2018IEEE,Aijaz2017radio,Aijaz2016towards} \\
& Data analysis and processing & \cite{Steinbach2011haptic,Gaodeep2016,Nozaki2018impedance,Hapticdatatransfer2016,Baran2016comparative} \\
\hline
& Opportunities, challenges, and enablers  & \cite{Bastug2017commun} \\
Wireless augmented/virtual reality & Resource management and scheduling & \cite{Chen2018virtual,Huang2018MAC} \\
& Cooperative communications & \cite{Ge2017multipath,Gu2018performance} \\
& eHealthcare scenarios & \cite{Avola2018toward,Bortone2018wearable} \\
\hline
Autonomous, intelligent and  & Automotive with V2X systems & \cite{Segata2015trans,Hung2017virtual,Di2017NoMA}  \\
cooperative mobility systems & Cooperative automated driving & \cite{Dressler2019cooperative,During2016cooperative,Segata2015trans} \\
& Internet of drones/UAVs & \cite{Kortelingtrans97,Shaikh2018robust,Mozaffari2017UAV,Motlagh2017} \\
& Mobile AR/VR & \cite{Wiederhold2018VR,Qiao2019Proc}
 \\
\hline
\end{tabular}
	\vspace{-15 pt}
\label{tab: TIsuecases}
\end{table*}

\subsection{Feasibility Analysis of Existing Wireless Technologies}
\label{sec:_sec21}
Since existing wireless technologies and protocols are mostly designed for the conventional HTC traffic, they are not sufficient to meet the stringent latency and reliability requirements of emerging TI applications which require M2M and M2H interactions \cite{Feng2017HCCA}. Therefore, it is crucial to analyze the feasibility of the existing technologies/protocols for TI application scenarios and to explore the adaptations of the existing protocols for TI applications. In this regard, some recent works have analyzed the feasibility of a few techniques for TI applications in different settings, which are briefly discussed below.

The factory automation can be considered as an example use case of TI, which has very strict requirements in terms of latency and reliability, i.e., end-to-end latency of about $1$ ms with a failure rate as low as $10^{-9}$ \cite{Yilmaz2015analysis}. In contrast to the conventional factory automation case which is mostly based on the wired connectivity, location flexibility of a large number of various devices in a factory can be enhanced with the robust and deterministic wireless connectivity. In this regard, authors in \cite{Yilmaz2015analysis} analyzed the feasibility of designing an Orthogonal Frequency Division Multiplexing (OFDM) based 5G radio interface for a mission critical MTC application. By considering a factory automation example, the authors showed the possibility of achieving sub-millisecond latency and high reliability with a very high availability of coverage while employing the considered OFDM-based 5G radio interface. Furthermore, considering the similar factory automation scenario, authors in \cite{Johansson2015radio} explored the viability of using wireless communications for low-latency and high-reliability communications. Via simulation results, it has been shown that by utilizing spatial diversity and short transmission intervals without retransmissions, it is possible to obtain very low error rates and delays over a wireless channel even in the scenarios with fast fading signals, interference, antenna correlation and channel estimation errors. Also, by considering a realistic factory environment, authors in \cite{Ashraf2016ultrareliable} evaluated the performance of LTE and 5G cellular systems by considering the connection of Base Station (BS) to the automated programmable logical controller with a negligible latency. Via numerical results, authors highlighted the need of new 5G radio interface for the applications with about $99.999$\% reliability and less than {1} ms end-to-end latency requirements.


Another TI application scenario is the video communications system, which is envisioned to support emerging cyber-physical networking, NCSs and dynamic interactive applications. In video communications, the end-to-end delay usually refers to
glass-to-glass delay, which is defined as the difference in time from  the photons of an event passing via the glass of a camera lens to the corresponding photons displayed on a monitor being passed through the display's glass \cite{Bachhuber2016system}. Current video systems suffer from a significantly large glass-to-glass delay of about $50$ ms to $400$ ms, which depends on the physical separation of two ends and how many networks the photons have to traverse. Through experimental results conducted in an Android platform, authors in \cite{Bachhuber2017videcomm} showed that current digital video applications are not yet suited for TI applications. Thus, there is a crucial need to lower the video communications delay in order to satisfy the requirements of TI applications such as tele-operation, autonomous connected driving and wireless AR/VR.

Due to time-varying channel fading and interference, providing reliable communications in wireless environments becomes challenging. To satisfy the stringent reliability requirements, one can study the signal quality outage performance by analyzing some suitable performance metrics such as Signal to Noise plus Interference (SINR) outage. In this regard, authors in \cite{Pocovi2015outage} performed the signal quality outage performance analysis of a cellular network and noted that the conventional Multiple-Input Multiple-Output (MIMO) systems with $2 \times 2$ or $4 \times 4$ antenna configurations are not sufficient to satisfy the stringent reliability requirements. To enhance this performance, the same authors suggested to utilize macroscopic diversity as well as interference management techniques along with MIMO schemes to enhance the SINR outage performance of a cellular network. However, while employing the considered MIMO diversity techniques, diversity-multiplexing tradeoff issue should be properly taken into account.

In terms of lab experimentation, authors in \cite{Piltz2016TI} demonstrated the implementation of a Software Defined Radio (SDR) based wireless broadband communication system  with $20$ MHz bandwidth and measured the basic performance indicators (throughput and delay) via the end-to-end measurements. Via experimental results, the authors showed the feasibility of real-time implementation of very low delay with a round-trip delay of about $1$ ms on an SDR-based platform. However, the experiment was carried out with a single wireless link in a basic setting without considering any interfering sources, and various delay components such as queueing delay and protocol delay were neglected.

Furthermore, authors in \cite{Gringoli2018WiFi} studied the feasibility of the existing IEEE 802.11 (Wi-Fi) technology for low-latency TI applications while considering both the IEEE 802.11 Direct-Sequence Spread Spectrum (DSSS) and IEEE 802.11g OFDM schemes. Mainly, the effect of imperfect synchronization resulting from the physically uncoupled devices utilized for simultaneous Wi-Fi transmissions, i.e., concurrent timing offset and concurrent frequency offset, were studied with the help of simulations by using SDR-based testbeds. The results indicated different behaviors of the IEEE 802.11 DSSS and IEEE 802.11 OFDM systems with regard to the aforementioned limiting factors resulted from concurrent transmissions. Also, the main parameter limits under which the IEEE 802.11 technology can support ultra low-latency TI applications were identified. Moreover, authors in \cite{Feng2017HCCA} investigated the potential of one of the latest MAC protocols of IEEE 802.11 networks, namely the Hybrid Coordination function controlled Channel Access (HCCA), towards supporting TI applications. Mainly, the wireless queuing latency from a tactile user to the access point has been analyzed for the IEEE 802.11 HCCA protocol, and its closed-form expression has been provided which can be used to select suitable HCCA parameters towards satisfying the desired latency requirement. Besides these MAC protocols, it is important to design load-aware MAC protocols for TI applications since the latency and reliability are very much dependent on the network traffic load.

One of the enabling platforms for haptic communications in Internet enabled eHealthCare applications is Wireless Body Area Networks (WBANs), where the on-body actuators/sensors can be used to support steering/control and haptic communications. The Smart Body Area Network (SmartBAN) proposed by European Telecommunication Standards Institute (ETSI) in 2015 \cite{ETSIsmartBAN} is considered as an important ultra-low power and low-complexity WBAN technology for health monitoring and reporting applications. In this regard, authors in \cite{Ruan2017towards} carried out the delay performance analysis of downlink transmission for SmartBANs with the objective of meeting the $1$ ms delay constraint required for TI applications. Also, the downlink delay of the SmartBAN system was modelled analytically and the impact of the selection of downlink transmission duration on the downlink delay and energy performance was studied. Via numerical results, it was shown that in contrast to the conventional exhaustive transmission, a fixed length transmission mechanism can provide less than $1$ ms delay required for TI Healthcare applications.

The TI applications depend on the interactions of human beings with the environment via our senses and these interaction times depend on the sensory stimulus as well as the state of whether the human is prepared for a particular situation or not \cite{Fettweis2014}. A faster reaction time is needed when a human is prepared for a scenario. The speed of this interaction is limited with our perceptual processes  and the technical requirements of TI systems required for real-time interactions  depend on the involved human senses. For example, the reaction time for human auditory sense is  about $100$ ms and modern telephone systems have been designed to transmit voice within $100$ ms in order not to feel the delayed conversation. Similarly, the typical reaction time for human visual sense is in the range of $10$ ms and modern TV sets  use the minimum picture refresh rate of about $100$ Hz (which translates to the maximum inter-picture delay of about $10$ ms)  for the users to experience seamless video. However, very rapid response time of about $1$ ms is required in the scenarios where speed is important for humans including a response from the visual display and the movement of heads while using VR goggles.

\section{Haptic Communications over Wireless TI}
\label{sec:_sec3}
Haptic communications provides an additional dimension over the conventional audiovisual communication to enable truly immersive steering and control in remote environments \cite{Aijaz2017tactile}. Although audio/visual information can provide the feeling of being present in the remote environment, it is not possible to have complete immersion without the exchange of haptic information in the form of various parameters such as force, motion, vibration and texture. The process of sending  command signals from the user to the remote environment and the reception of the haptic feedback signals from the remote environment form a closed control loop in a haptic communication system \cite{Aijaz2018IEEE}.

The haptic communications using networked tele-operation systems is one of the important application areas of the TI, which demands for an efficient and timely exchange of  kinesthetic or tactile information. In contrast to the communications of audio and video signals, haptic signals in the bilateral tele-operation systems are bidirectionally exchanged over the networks, thus creating a global control loop between the human users and the teleoperators/actuators \cite{Antonakoglouhaptic}. The haptic communications and TI have the service and medium relationship similar to the relation between voice over Internet protocol and the Internet \cite{Aijaz2017tactile}.

As depicted earlier in Fig. \ref{fig: TIarchi}, a haptic communication system usually consists of a master device (user) at one end, a communication network in between, and the slave device (tele-operator) at the other end. The master device sends the position and/or velocity data while the slave device transmits the haptic feedback data (surface texture, force/position) along with the audio and video data streams via the underlying communication channel \cite{Antonakoglouhaptic}.

The haptic sense, i.e., the sense of touch establishes the links between the unknown environments and humans similar to audio and visual senses. The main difference from the audio and visual sense is that the haptic sense occurs bilaterally. The haptic feedback is usually of two types: (i) kinesthetic feedback which provides information about the velocity, force, position and torque, and (ii) tactile feedback which provides information about the texture, surface and friction \cite{Aijaz2018IEEE}.

\subsection{Key Challenges}
\label{sec:_sec31}
The main challenges for realizing haptic communications over the wireless TI include the following \cite{Challengeshapticcommunication,Aijaz2017tactile,Antonakoglou2018haptic}.

\begin{enumerate}
\item \textbf{Ultra-High Reliability}: One of the crucial requirements of TI applications over wireless networks is ultra-high reliability, which can be impacted due to a number of factors in wireless networks such as the lack of resources, uncontrollable interference, reduced signal strength and equipment failure \cite{PopovskiURLLC}. Although the conventional video and voice applications naturally allow the graceful degradation of service quality in worse communication situations, this is not applicable for haptic applications due to the possible unstable situation which may result from the delayed arrival or the loss of critical haptic information \cite{Aijaz2017tactile}. Also, various layers of the conventional protocol stack including the MAC layer, transport layer and session layer need to be revisited to enable ultra-high reliability in haptic communications. For example, in the transport layer, the Transport Control Protocol (TCP) provides high reliability at the expense of a very high protocol and high latency while the user datagram protocol can provide low protocol overhead but at the cost of reliability \cite{Challengeshapticcommunication,Antonakoglou2018haptic}. Also, maintaining a small packet header to the payload ratio is one of the main issues since the payload of haptic data packet for 3-DoF (Degrees of Freedom) haptic system is only 6 bytes, but the header size in IPV6 version is 40 bytes \cite{Aijaz2017tactile}.

\item \textbf{Ultra-Responsive Connectivity}: There are several challenges in achieving about $1$ ms end-to-end latency required by many TI applications. All the factors contributing to this end-to-end latency across different protocols-layers, air interface, backhaul, hardware and core Internet, should be optimized to meet this target. For example, in the physical layer, shorter Transmission Time Interval (TTI) is highly desirable to lower the over-the-air latency, but this will require higher available bandwidth. After investigating optimum combinations, the achievable end-to-end latency is limited by the finite speed of light, which sets an upper bound for the maximum separation between the tactile ends.

\item \textbf{Efficient Resource Allocation}: The radio allocation in wireless/cellular networks becomes challenging with the incorporation of haptic communications since the available resources need to be shared between haptic communications and Human-to-Human (H2H) communications or machine-type communications, which have diverse and conflicting service requirements. Among these systems, haptic communications should get priority for resources due to its stringent QoS requirements. Also, due to the bidirectional nature of haptic communications, symmetric resource allocation with the guarantee of minimum constant rate in both the uplink and the downlink is required. Furthermore, flexible resource allocation techniques across different protocols layers including adaptive flow management and network slicing \cite{Grasso2018design} with the on-demand functionality needs to be investigated to enable the coexistence of haptic communications with other systems.

\item \textbf{Multi-Modal Sensory Information}: TI systems need to consider audio and visual feedback in addition to the haptic feedback in order to have the increased perceptual performance. The main challenge here is the cross-modal asynchrony, which may arise due to the heterogeneous requirements of haptic, visual and auditory modalities in terms of various parameters such as latency, transmission rate and sampling rate. This leads to the need of effective multiplexing scheme which can integrate these different modalities by exploiting their priorities. Although some works \cite{Eidadaptivemux,CizmeciJune2014} in the literature have suggested application layer-based multiplexing schemes, further research on exploring effective multiplexing schemes across different protocol layers for integrating the aforementioned modalities in dynamically varying wireless environments is needed.

\item \textbf{Collaborative Multi-User Haptic Communications}: In multiuser haptic scenarios, multiple users need to interact and collaborate in a shared remote environment, thus requiring the formation of a peer-to-peer overlay to enable the collaboration among multiple users \cite{Keong2015survey}. This overlay creation step raises additional challenges in terms of meeting the requirements of TI applications since the overlay routing and IP-level routing may further increase the end-to-end latency \cite{Aijaz2017tactile}.

\item \textbf{Area-Based Sensing and Actuation}: In contrast to the most existing haptic devices with the single-point contact for the tactile and kinesthetic feedbacks, haptic communications need to deal with the touch-based sensations across the surfaces including the palm of the hand or other parts of the body. For this purpose, area-based or distributed sensing and actuation needs to be incorporated in haptic devices \cite{Aijaz2017tactile}. Also, this will significantly impact the conventional communication requirements due to the increased data rates and a different perception for the case of a data loss.

\item \textbf{Definition of Novel Performance Metrics}: A common approach to  evaluate and compare the performance of different haptic communications systems is still not available and it is important to investigate suitable performance metrics towards analyzing and comparing the performance of various haptic systems over the TI \cite{Sachs2019adaptive}. A suitable mapping between the QoS metrics delivered by the involved communication networks and haptic data processing techniques (data reduction/control) is necessary for the development of novel joint optimization techniques for communications, compression and control. Some key performance metrics to evaluate the performance of haptic communications over TI include Quality-of-Experience (QoE) and Quality-of-Task (QoT). Out of these metrics, QoE mostly refers to the difference of the physical interaction across a network and the same manipulation carried out locally, thus defining the transparency of the system \cite{Xu2016energy}. On the other hand, QoT measures the accuracy by which a tactile user can perform a particular task \cite{Sachs2019adaptive}.

\item \textbf{Stability for Haptic Control}: The global control loop in haptic communication involves different entities including the human, the communication network, the remote environment, and the energy exchange among these components takes place via various commands and feedback signals \cite{Aijaz2017tactile}. However, in practical wireless environments, the system instability may arise  due to time varying delays and packet losses. Therefore, one of the main challenges in wireless TI systems is to maintain the stability of the control loop system since the instability may lead to a significant degradation of the impressiveness to the remote environment.

\item \textbf{Haptic Codecs}: In order to perform the digitization of haptic information, haptic signals are usually sampled at the rate of 1 kHz, generating $1000$ packets per second \cite{Challengeshapticcommunication}. Although different haptic data compression techniques have been investigated by utilizing the limits of human haptic communications, the main challenge in the context of TI applications is to develop a standard family of haptic codecs by incorporating both the kinesthetic and tactile information. Also, the developed haptic codes should be able to perform effectively in time-varying wireless environments.

\item \textbf{Haptic Sensors and Actuators}: Haptic devices such as haptic sensors and haptic actuators are important components of haptic communications systems. Among these devices, haptic sensors sense the tactile information by interacting with the surroundings and they are usually mounted at the tele-operator end. This sensed information is then relayed back and transmitted to the end-user in the form of the haptic feedback by the haptic actuators (also called haptic feedback devices). In the following, we briefly describe the functionalities of these devices and the associated issues \cite{Challengeshapticcommunication}.

Haptic sensors are mainly pressure sensors to detect the underlying pressure. They mostly follow the capacitive and resistive methods. The capacitive sensor is made up of a flexible dielectric in between two conducting plates and the change in the capacitance occurs due to a reduction in the distance between these two plates caused by the pressure to be measured. On the other hand, a resistive method uses a pressure-sensitive resistive material whose resistance changes upon applying the pressure. This resulted pressure in both cases can then be measured with the help of an electric circuit and digitized with an analog to digital converter. In terms of the underlying materials, the capacitive haptic sensors mostly use elastomer polydimethylsiloxane (also called silicone) as a dielectric, while the resistive haptic sensors commonly use pressure sensitive piezo-resistive materials.

The main issues associated with the capacitative sensors include non-idealities and sensitivity. Furthermore, regardless of the type of the haptic devices, several parameters such as range, sensitivity, response time, spatial resolution, cost, temperature dependance and complexity need to be taken into account when selecting a haptic sensor; and the importance of the aforementioned parameters depends on the employed use-case. Another issue is the placement strategy for single-element sensors and sensor arrays. With the increase in the array size, the sensing time and the overall power consumption also increase. Furthermore, another challenge is the design of sensing circuits having high spatial resolution. In summary, the optimization of the spatial resolution, scan time, sensitivity, and placement is important for haptic sensors.

Haptic actuators can be of type cutaneous or kinaesthetic. The cutaneous type deals with the muscle or force tension and the sense of the relative position of the neighbouring body parts. On the other hand, the kinaesthetic type mainly deals with the touch related to the skin, including vibration, pressure, pain and temperature. In haptic feedback systems over wireless TI, it is crucial to provide a feeling of touch to the user which is similar to the feeling the user would receive in the real-world context. One of the main current challenges is to design a practical and lightweight haptic display, which have the capabilities of both the cutaneous and kinaesthetic feedback.
\end{enumerate}
\subsection{Requirements and Potential Enablers}
\label{sec:_sec32}
Towards supporting the haptic communications in LTE-A based cellular networks, authors in \cite{Aijaz2016towards} identified the key requirements of haptic communications from the perspective of radio resource allocation and studied the radio resource allocation problem in LTE-A based networks. Due to the presence of a global control loop and the bidirectional nature, haptic communication has unique requirements for radio resource allocation in LTE-A based cellular networks in contrast to the radio resource allocation for voice/video communications. These unique requirements are listed in the following \cite{Aijaz2016towards,Aijaz2018IEEE}.
\begin{enumerate}
\item Symmetric resource allocation in both the uplink and the downlink is required due to the involvement of the bidirectional flow of feedback and command signals between the master domain (haptic device end) and the slave domain (Teleoperator end).
\item Radio resources in the conventional multimedia communications are usually allocated independently in the uplink and downlink. This independent resource allocation approach is not suitable for haptic communications because of the involved coupling of uplink and downlink resulted due to the bidirectional information exchange. The QoS degradation in the forward path of haptic communications directly affects the QoS of the reverse path \cite{Steinbach2012haptic}, thus resulting in the need of joint resource allocation in both the uplink and downlink.
\item The requirement of bounded delay needs to be considered in the resource allocation problem in order to ensure the stability of the global control loop.
\item It is essential to guarantee the minimum rate throughout a complete haptic session.
\end{enumerate}

In the following, we review the related works which discussed the issues and potential enablers of haptic communications.

Authors in \cite{Aijaz2016towards} translated the aforementioned radio resource requirements into a power and Resource Block (RB) allocation problem considering the constraints of the uplink and downlink multiple access schemes. The formulated problem was decomposed by using the optimal power control policy, and then was transformed into a binary integer programming problem for the RB allocation. Also, the authors employed a low-complexity heuristic algorithm for joint scheduling in both the uplink and downlink by considering the unique requirements of haptic communications.

Future 5G and beyond networks are expected to support haptic communications, leading to the concept of human-in-the-loop mobile networks in contrast to the past wired human-in-the-loop  solutions \cite{Aijaz2018IEEE}. However, to enable the real-time haptic interaction over wireless networks, current wireless technologies need to be improved to satisfy the stringent communication requirements in terms of very low latency and ultra-high reliability. In this regard, authors in \cite{Aijaz2018IEEE} discussed various architectural aspects and radio resource allocation requirements of the human-in-the-loop mobile networks.

Radio resource slicing can be considered as one of the promising solutions to provide flexible resource allocation in haptic communications over 5G and beyond networks. However, most of the existing radio resource slicing solutions are unable to deal with the varying traffic demands and channel conditions of highly dynamic wireless environments, where each network slice follows different dynamics and requires an independent optimal slicing period (the period after which the slice-size needs to be recalculated) \cite{Aijaz2017radio}. In order to achieve the maximum utilization of the limited radio resources in these environments, the slicing strategy should be as dynamic as possible in contrast to the conventional static slice allocation.

In the above context,  authors in \cite{Aijaz2017radio} proposed a radio resource slicing framework, named as Hap-SliceR, which defines the  network-wide optimum radio resource allocation strategy for haptic communications over 5G networks. The proposed strategy flexibly allocates the radio resources to different slices while taking the dynamics and utility requirements of different slices into account. A Reinforcement Learning (RL) based technique was utilized which enables the learning entity  (responsible for slicing) to not only learn the size of each slice, but update it dynamically based on the underlying requirements. Furthermore, a post-decision state learning approach \cite{Mastronarde2011} was employed on the top of the RL approach to further enhance the efficiency of the slicing strategy.

In order to accurately predict haptic properties and interact with the world, humans usually combine visual predictions and feedback from the physical interactions. In this regard, authors in  \cite{Gaodeep2016} proposed a method to classify surfaces with haptic adjectives by utilizing both the physical and visual interaction data and demonstrated that more accurate haptic classification can be obtained with both the visual and physical interaction signals.

Haptic communication needs to deal with the rapid transfer of a large amount of haptic data due to a short sampling period and high frequency communication requirements. The bilateral control in haptic communication usually involves the following types of information flow: bidirectional communication of angle, angular velocity, and torque \cite{Natori2010timedelay}. With the growing number of haptic devices, the limited capacity of the communication channel becomes one crucial aspect to be addressed. In this regard, investigating suitable traffic reduction techniques is one of the important research issues. To address this, authors in \cite{Nozaki2018impedance} recently proposed an impedance control method for the traffic reduction of haptic data by utilizing the equivalency between the impedance field and the standard bilateral control. This equivalency was validated theoretically, and via simulation and experimental results, it was shown that the transmission of only the force information can provide similar performance to that of the standard bilateral control.

Towards meeting the necessary QoS requirements of haptic data transfer over the Internet, the considered transport protocol should possess several qualitative features including reliability, prioritization, congestion/flow control, buffer optimization, differential coding, packetization and synchronization \cite{Kokkonis2016network}. In this regard, authors in \cite{Hapticdatatransfer2016} discussed the transmission requirements of the haptic data and the necessary QoS for the haptic data with the objective of maximizing the quality of experience for the haptic users. With the help of experiments, the authors demonstrated the feasibility of haptic data transfer over the Internet subject to some limitations in terms of geographical distance and network conditions.

One of the crucial issues in the area of bilateral teleoperation is the delay involved with the communication channel between the master and slave operators \cite{Tian2012wireless}, which may result in performance degradation and system instability.
This delay based performance degradation is mainly related to the bandwidth and throughput trade-offs in the NCSs. To fulfill the requirements of stability and better precision, the NCSs usually demand for much higher sampling frequencies than those needed for the high performance communication network \cite{Nakano2015quantization}, thus the need of two different Nyquist frequencies (i.e., high speed sampling for the controller and low speed sampling for the network) \cite{Baran2016comparative}.

Haptic communications can be broadly classified into open-loop and closed-loop haptic communications, which have very different communications requirements \cite{Sachs2019adaptive}. The exchange of kinesthetic information in bilateral teleoperation systems with the kinesthetic feedback can be considered as an example of open-loop haptic communications while haptic perceptual systems such as surface haptics can be considered as a use-case example of closed-loop haptic communications. The closed-loop haptic communications system needs to meet strict latency constraints due to the requirement of sending new force or sensor readings immediately while the open-loop haptic systems do not suffer from these stringent delay requirements since they do not involve a global control loop.

\subsection{Use Case Example: Teleoperation Systems}
\label{sec:_sec33}
Authors in \cite{Antonakoglouhaptic} provided a detailed survey of haptic communications in the networked tele-operation systems over the TI by considering the remote environments of low and intermediate dynamics. They investigated the following three main domains: (i) reliable and low-latency communication network, (ii) intelligent data processing, and (iii) stability control methods. In the following, we briefly discuss some aspects related to haptic communications over the teleoperation systems.

A multi-modal teleoperation (telepresence and teleaction) system, also called telephatic system, usually comprises three main entities: (i) one human operator with a haptic interface (master device), (ii) a communication channel, and (iii) one teleoperator (slave actuator) \cite{Antonakoglouhaptic}. The main objective of this system is to provide human users with multimodal feedback including visual, auditory, and haptic feedback  with the continuous development in the involved hardware and software. Depending on the involved communication delays and interaction levels that an end-user may experience, tele-operation systems in general can be categorized into direct control and supervisory control. The direct control system can be further divided into closed-loop with the negligible delay and the time-delayed closed loop, while the supervisory control can be either autonomously or semi-autonomously controlled.

From the communication perspective, there are mainly the following two challenges for the effective design and operation of teleoperation systems. The first challenge is that communicating kinesthetic information between the master and slave ends in a teleoperation system requires a high packet transmission rate of about $1000$ or more haptic data packets per second, leading to an inefficient data communication due to the depletion of the network resources. To address this issue, suitable haptic data reduction and compression techniques for teleoperation systems need to be investigated. The second challenge is that teleoperation systems are very sensitive to the latency and data loss \cite{Lawrence93rans}. Furthermore, authors in \cite{Steinbach2012haptic} analyzed the force-velocity control architecture of a telepresence system for the positions of human system interface and teleoperator with and without considering the involved communication delay. They reported that even a small amount of delay or packet loss rate can cause stability issues in a bilateral teleoperation system, leading to the service quality degradation.

In many areas of telepresence and teleaction systems including tele-education, tele-manipulation in dangerous environments, teleconferences, telerobotics, minimally invasive surgery and on-orbit teleservicing, a realistic presentation of the remote environment is crucial to enhance the performance of haptic transmission as well as to guarantee the QoE to the haptic user. One of the important approaches to enhance the high quality interaction in immersive communication is to employ haptic data reduction with the help of haptic data compression. In principle, the compression of haptic signals is different than the compression of audio and video signals due to the strict stability and delay requirements. In this regard, authors in \cite{Steinbach2011haptic} provided a comprehensive discussion on a number of haptic data reduction methods including perceptual deadband-based haptic data reduction, velocity-adaptive perceptual deadband-based, signal-based predictive coding, model-based predictive coding and event-based coding of haptics.

\section{Wireless Augmented and Virtual Reality}
\label{sec:_sec4}
The emerging Virtual Reality (VR) provides an immersive environment for the user to interact with the world by means of a head-mounted display while the Augmented Reality (AR) enables the user to visualize the superimposing content over the real world in mobile devices including smart phones and laptops. These VR and AR applications demand for real-time interactions and the flow of massive information, and thus will introduce new design challenges in future networks in terms of improving several performance metrics such as network throughput, delay performance and capacity \cite{Ge2017multipath}.

Wireless VR applications need to concurrently support ultra-high data-rate, ultra-high responsive speed and ultra-high reliability \cite{Huang2018MAC}. AR and VR have a significant number of emerging applications such as high mobility using automotive, high-capacity upload from the event venues, high-bandwidth 6 DoF experiences, and low-latency remote control, and the TI \cite{QualcomAR2018}.
However, today's powerful VR prototypes are mainly based on wired/cable connection since the high resolution video at high frame-rates cannot be achieved using today's wireless technologies and a perfect user interface is still lacking \cite{Bastug2017commun}. Also, current VR applications are mostly based on wired connections with the limited actions that the user can take and it is very important to deploy VR services in the wireless environment to have truely immersive VR applications.  However, there are several issues regarding the transmission of VR services in a wireless environment, i.e., in terms of tracking accuracy, data-rate and latency \cite{Bastug2017commun}. To address these issues, there are several research attempts in the direction of realizing interconnected VR, which are reviewed in the following.

The main objective of the VR is to create a digital real-time experience that reflects the full resolution of the human perception, i.e., regenerating every photons that our eyes observe, including small vibrations that our ears can detect, as well as other cognitive aspects such as smell and touch. In this regard, the authors in \cite{Bastug2017commun} highlighted the current and future trends of VR systems towards reaching a fully interconnected VR society, and discussed scientific challenges and the key enablers for total interconnected VR systems.

Furthermore, the transfer of VR video streaming needs to satisfy the following three main requirements: (i) ultra-high data rate, (ii) ultra-high responsive speed, and (iii) ultra-high transfer reliability, which are also called three Ultra-High (3UH) requirements. The first requirement of ultra-high data rate arises from the $360^{\circ}$ view-port and high-definition video quality while the second requirement  of ultra-high responsive speed arises from the low motion-to-photon latency request (normally $10$-$20$ milliseconds). Similarly, the third requirement necessitates the need of guaranteeing a satisfactory Quality of Experience (QoE).

In the context of multiple VR users, one of the main challenges is to concurrently support the maximum number of VR users while guaranteeing their 3UH QoE, thus leading to the need of designing efficient scheduling techniques. In this regard, authors in \cite{Huang2018MAC} proposed a multiuser MAC scheduling scheme for the VR service in MIMO-OFDM based 5G systems, which comprises three main functions, namely: video frame differentiation and delay-based weight calculation, link adaptation with dynamic Block-Error-Rate (BLER) target and the maximum aggregate delay-capacity utility based spatial-frequency user selection.

With regard to supporting VR applications with cellular networks, authors in \cite{Chen2018virtual} studied the resource management problem for a small-cell based wireless network having wireless VR users, and proposed a multi-attribute utility theory based VR model by jointly taking account different VR metrics including processing delay, tracking accuracy and transmission delay. In the considered scenario, the small-cell BSs acting as VR control centers gather the tracking information from the VR users over wireless links in the uplink, and subsequently send the 3D images and the associated audio information to the VR users in the downlink.

Due to the need of real-time interactions and the flow of massive information in VR/AR applications, future networks need to address several new challenges mainly in terms of meeting the network capacity demand and delay performance. In contrast to the transmission of the traditional video applications, VR/AR applications have very strict uplink/dowlink latency requirements. For example, for the transmission of $360^{\circ}$ video in a VR application, Motion-to-Photon (MTP) latency should be within $20$ milliseconds in order for the users not to feel dizzy. Towards meeting these requirements of low latency and reliable data transmission in VR/AR applications, authors in \cite{Ge2017multipath} recently proposed to utilize an SDN architecture for 5G small-cell networks, and subsequently utilized a multi-path cooperative route scheme for fast wireless transmissions to the desired user from multiple edge data-centers.

Furthermore, authors in \cite{Gu2018performance} carried out the performance analysis of interconnected VR systems by utilizing a stochastic geometry tool in a 3D urban ultra-dense scenario consisting of densely deployed low-power nodes. Subsequently, a two-stage cooperative transmissions scheme was proposed to mitigate the harmful inter-cell interference in the considered ultra-dense scenario, and the expressions for average VR data-rate and latency probability were derived. The accuracy of the derived expressions and the practicability of the proposed two-stage cooperation scheme was demonstrated via numerical results.


Moreover, the generation of VR video requires the knowledge of tracking information related to the interactions of the VR users with the underlying VR environment. The tracking delay as well as the tracking accuracy will significantly impact the generation and transmission of VR videos, thus degrading the users' QoE. The tracking delay is a crucial aspect to be considered in VR video transmission in contrast to the conventional static High Definition (HD) videos \cite{Chen2018virtual}. Furthermore, compared to the HD video where only the video transmission delay is the main concern, several parameters such as video transmission delay, tracking information transmission delay and the delay occurred in generating VR videos based on the tracking information need to be jointly considered in VR video applications.

The resulting immersive experiences from the wireless VR/AR will facilitate the new way of communicating, working and entertainment in the next IoT era. Mainly, there are four types of VR Head Mounted Device (HMD): (i) PC VR (examples: Oculus Rift and HTC Vive), (ii) Console VR (example: playstation VR), (iii) mobile VR (example: Samsung gean VR, Google Daydream), and (iv) wireless all-in-one HMD device (example: Intel Alloy) \cite{Hou2017wireless}. Currently, these devices are heavy and large, and industries are investigating the ways to make them lighter and portable. In most of today's HMDs, the main processing happens locally, and therefore, this raises the portability and mobility issues. To address these issues, authors in \cite{Hou2017wireless} proposed to carry out rendering either at the cloud-servers or at the edge-servers in order to enable the mobility and portability of the VR experiences. They also reported on the advantages and disadvantages of rendering at the cloud-servers, edge-servers and local devices.

In terms of emerging applications, the VR is enabling important use-cases in the field of medical surgery via immersive VR and human-computer interaction. Recent technological advances in imaging techniques for the visualization of organs/tissues and minimally invasive instruments such as endoscope are leading to the possibility of immersive VR in medical surgery applications \cite{Avola2018toward}. One of the main issues in implementing VR in surgery applications is to integrate various technologies to facilitate surgeon in exploring and treating a target organ/tissue in a way that they were inside the patient. In this regard, authors in \cite{Avola2018toward} investigated a 3D immersive VR-based endoscopic system to support minimally invasive interventions and intra-body cavities, and validated the significance of the proposed method with the support from the skilled surgeons.


Another emerging application of VR is rehabilitation treatment in the clinical practice. In this regard, authors in \cite{Bortone2018wearable} proposed an immersive VR rehabilitation training system with wearable haptics towards enhancing the engagement of child patients with neuromotor impairments. Also, an experimental rehabilitation session was conducted with children, who suffered from celebral palsy and development dyspraxia, and a performance comparison of children and adult control groups was carried out.


\subsection{Key Technical Challenges}
\label{sec:_sec41}
The main technical challenges for interconnected wireless VR \cite{Bastug2017commun} are briefly discussed below.
\begin{enumerate}
\item \textbf{Novel Information Theoretic Principles}: The investigation of basic information theoretic principles such as Shannon theory for the wireless VR scenarios is one of the important future research directions. Under this theme, different enablers such as coding design to minimize the feedback delay, haptic code design for VR systems, source compression under the imperfect knowledge of input distribution, source coding with the complete knowledge and  compressed sensing with imperfect structural knowledge, need to be investigated for wireless VR scenarios.
\item \textbf{Quality-Rate-Latency Trade-off}: Another important issue is to investigate the quality level per content which maximizes the quality of immersive VR experience under the given network topology, storage and communication constraints. Furthermore, finding the optimal payload size for a given content to maximize the information transfer rate under the given latency and rate constraints for the self-driving car scenario is another interesting optimization problem.
\item \textbf{Scalability and Heterogeneity}: Another challenge is to analyze the very large VR systems and networks having diverse views and raw information.
\item \textbf{Localization and Tracking Accuracy}: The accuracy of the existing localization and tracking techniques needs to be enhanced to achieve a fully immersive VR experience in different scenarios such as human eyes tracking and object localization.
\item \textbf{Energy-efficient VR}: To achieve the green VR, power consumption in terms of communication, computing and storage should be minimized as much as possible for a given target VR user's immersive experience.
\item \textbf{In-Network versus In-VR Computation}: Another issue is to find out which level of task/computation should take place in the VR headset and which component on the network-side depending on bandwidth-latency-cost-reliability trade-offs.
\item \textbf{Quantum Computing}: The emerging quantum computing paradigm \cite{Nawaz2019IEEE} can address different issues related to computation in VR systems by exploiting entanglement and superposition principles and also by bridging the virtual and physical worlds.
\item \textbf{Privacy Preservation}: It is a crucial challenge to investigate intelligent mechanisms which can automatically preserve privacy without providing burden on the user-side/haptic devices.
\end{enumerate}

\subsection{Potential Enablers}
\label{sec:_sec42}
The key enablers for the interconnected wireless VR include the following \cite{Bastug2017commun}.
\begin{enumerate}
\item \textbf{Caching/Storage/Memory}: Caching/storage in different entities of the VR systems will have a crucial role  since it is not efficient to access the content-server each time the request arises. Various kinds of side knowledge such as the user's location, mobility patterns and social ties can be exploited to make the decision on which contents to cache and where to cache them.
\item \textbf{Collaborative Local/Fog/Edge and Cloud Computing/Processing}: Due to the huge computational capacity and storage capacity available at the cloud, it becomes highly beneficial to process computationally demanding tasks at the cloud-center. However, this cloud-side processing is not suitable for the applications demanding low-latency and high QoS \cite{SKSIEEE2017}. On the other hand, edge/fog computing at the devices/aggregators/access points can support the applications/services demanding low-latency, location-awareness, high mobility and high QoS \cite{Alnoman2019emerging}. In this regard, collaborative edge/cloud processing \cite{SKSIEEE2017}, which can utilize the benefits of both the computing/processing paradigms, seems one promising enabler for wireless VR applications.
\item \textbf{Short-Range Wireless Communications}: Relevant contextual information can be shared among collocated VR uses by means of M2M or Device-to-Device (D2D) communications to address the network congestion issue.
\item \textbf{Context-Awareness and Analytics}: The acquisition of contextual information and the analysis of the acquired information to extract the meaningful information, play a crucial role in optimizing complex wireless networks including wireless TI. The contextual information may be in-device and in-network side information such as user location, velocity, employed MAC schemes and battery level.
\item \textbf{User's Behavioral Data and Social VR}: Other aspects to be analyzed towards enhancing the user's QoE are the user's behavioral data and social interactions. For example, novel solutions to address the issue of switching among multiple screens should be developed for wireless VR applications by utilizing a common data-driven platform.
\end{enumerate}


\section{Autonomous, Intelligent and Cooperative Mobility Systems}
\label{sec:_sec5}
In this section, we provide some discussions on TI applications in vehicular communications and unmanned autonomous systems including automotive with V2X communications, cooperative automated driving, Internet of drones and mobile AR/VR.

\subsubsection{Automotive with V2X Communications}
\label{sec:_sec51}
V2V communication is considered as one of the enabling technologies for low-latency communications among vehicles \cite{Schwarz2017signal}. However, a pure ad-hoc network architecture may not achieve the desired latency requirement due to scalability issues, and thus Vehicle-to-Infrastructure (V2I) communication is needed not only to improve the communication reliability, but also to reduce the reaction time of the vehicles. In this regard, authors in \cite{Hung2017virtual} proposed a time dynamic optimization approach to deal with the highly dynamic environment of vehicular networks (involving both the V2V and V2I links) by employing a virtualization technique. Two important rules to establish virtual heterogeneous networks include the complete isolation among different virtual networks and services and additional control signalling overhead of the proposed scheme. The proposed virtualization approach in \cite{Hung2017virtual} studied the optimization of the spectrum resources by considering the constraints on the network switching rate. Via the analytical results, the tradeoff between the network switching rate and the latency performance was analyzed.

The safety critical services play a crucial role in enhancing the driving experience via several emerging applications.
However, the main challenge in the vehicular networks is to achieve the desired low latency and high reliability of V2X services. Furthermore, since the existing LTE-based vehicular networks are based on the orthogonal multiple access technique and do not fully utilize the limited frequency resources, this may lead to the problem of data congestion and lower access efficiency in dense vehicular networks. In this direction, authors in \cite{Di2017NoMA} proposed a NOMA-based hybrid centralized/decentralized scheme towards reducing the access collision and enhancing the reliability of the network.

Furthermore, the V2I and V2I communications are important to future cars for collaborative autonomous driving and in-car entertainments \cite{5Gppautomative}. In these applications, there arises a need of transferring huge amount of information with a very low latency and the desired latency depends on the dynamics of the surrounding environments. Moreover, very high quality of audio and video information transfer is required in artificial intelligence-assisted collaborative autonomous driving. In addition, in remote driving applications, the exchange of haptic information requires very high degrees of freedom (in the order of several tens to hundreds). The achievable latency in the existing V2V and V2I links is more than $10$ ms, and this is not acceptable for handling autonomous vehicle traffic \cite{Cshe2016VTC,Gholipoor2018}.

Future V2V and V2I communications require very reliable connectivity to handle life-critical situations and the sensing data needs to be communicated to the drivers in real-time to make reliable and timely decisions, thus leading to ultra-low latency requirements and automotive high speed networks. In this direction, some of the recent trends  for supporting drivers in life-critical conditions include high-resolution cameras and sensors having high data-rate volume. Furthermore, automotive audio-video bridging and time sensitive networks are in the process of standardization \cite{Holland2019IEEE}. Also, emerging haptic applications in the automotive sector include the remote driving support of trucks, shuttles and road machines in the regions where it is difficult to maintain or serve, and where remote driving requires immediate feedback in order to make reliable decisions in life-threatening cases.


\subsubsection{Cooperative Automated Driving}
\label{sec:_sec52}
One promising application scenario of the TI is the cooperative automated driving, which is composed of two fundamental domains, i.e., vehicular networking and automated driving. In this TI application, one of the crucial  scenarios is cooperative adaptive cruise control or platooning since it demands for strict communication requirements in terms of high reliability and low latency, in order to ensure the safety gap of less than 5 m between the cars.  The message rates in the order of about 10 kHz is needed for large networks. This is beyond the capacities of existing vehicular networking technologies under the critical network congestion and unreliable wireless links. In this regard, authors in \cite{Dressler2019cooperative} discussed the future opportunities of utilizing TI principles by integrating multidisciplinary technologies from the mechanical engineering, control theory and communication protocol design. Furthermore, future research challenges for the realization of cooperative automated driving in terms of scalability and dependability, security and privacy, and public acceptance, have also been discussed.

The problem of scalability needs to be addressed from the perspective of resource allocation rather than the congestion resolution approach since this TI application may not be able to wait for the congestion to be resolved. For ensuring the full dependability of the automated driving system, the underlying TI communications need to be completely coordinated  by means of different physical communication channels (cellular, short-range radio techniques, mmWave and visible light communications) and they should be as uncorrelated as possible. Moreover, ensuring security is a critical issue since it is directly related to the public safety. From the communication perspective, there may occur two major types of attacks in the TI, namely, tampering of the onboard unit and exploiting radio for sending fake information but authenticated messages \cite{Dressler2019cooperative}. The main challenge in both of these attack scenarios is to have the timely transmission of reliable information required for cooperation. To address this, multiple communications links should be utilized in parallel for redundant information, and also the data should be combined and checked for consistency before feeding to the control algorithms.

Cooperative automated driving envisions the collaborative operation of a number of self-driving vehicles. However, most of the self-driving vehicles are based on single-vehicle control/sensing operations; moreover, the perception field of the vehicle is limited to the local coverage of onboard sensors. In order to concurrently ensure the safety and traffic efficiency, single-vehicle based perception/control needs to be completely transformed to multi-vehicle perception/control. In this regard, TI can act as a potential paradigm by enabling fast and reliable transfer of sensor data along with the haptic information related to driving trajectories among the vehicles via V2V/V2I or vehicle-to-any (V2X) communications. This will subsequently enable the functionalities of cooperative perception and manuevering \cite{During2016cooperative} for cooperative driving applications. Also, TI can enable the extension of the sensing range of the vehicle as well as the time horizon for the prediction of underlying situations for ensuring the safety of autonomous vehicles.

Platooning is considered as a potential approach to enhance the traffic flow efficiency and road traffic safety and it needs to deal with the strict communication requirements in terms of update frequency and reliability. In this regard, authors in \cite{Segata2015trans} exploited different communication strategies by considering the synchronized communication, transmit power adaptation and the requirements of the controller. Through simulation results, the authors demonstrated the potential of the combination of synchronized communication slots with transmit power adaptation for cooperative driving applications.

\subsubsection{Internet of Drones}
\label{sec:_sec53}
Another promising application of  TI is Internet of drones. Due to recent advances in control and communication technologies, future unmanned aerial vehicles, i.e., drones are expected to extensively deliver packages or emergency items such as medicine/medical equipment for patients and other urgent components. In this regard, big industrial players including Google, Amazon and DHL have already tested the delivery feasibility via drones \cite{Holland2019IEEE}. However, during these tests, only a few number of drones has been utilized, but the management of traffic for delivery drones becomes highly essential in the future by considering a huge number of drones owned by several companies. In addition, low-latency communication and dynamic route selection are crucial to avoid drone collisions and any damages underneath. In this regard, TI could be a promising paradigm for enabling ultra-high reliability, ultra-low latency and overall safety of future drone delivery systems. In addition to the transmission of real-time location, audio and video data, haptic (tactile and kinaesthetic) information can be transmitted over the underlying communication network to enhance the performance of future drones/UAVs.

\subsubsection{Mobile AR/VR}
\label{sec:_sec54}
VR has been considered as an effective tool to reduce the procedural pain  and psychological in the patients undergoing surgery with its analgesic effects of reducing respiration rate, heart rate, blood pressure and oral secretion. With the recent advances in cheaper, lightweight, and easy-to-use mobile technologies, there has been increasing interest in employing VR technologies in tele-health and home health-care scenarios, leading to the concept of mobile VR \cite{Wiederhold2018VR}. Besides validating the palliative effects of VR systems in clinical scenarios, there has been studies to analyze the efficacy of mobile phones in delivering VR services to the patients \cite{Mosso09,Wiederhold2018VR}. Although not as effective like the costly wired HMDs, mobile devices are shown to mitigate the pain-level and psychological indicators such as heart rate \cite{Wiederhold2014mobile}. Such mobile VR approach can enable the paradigm shift of moving palliative, rehabilitative and preventive care at the remote distances from the patients' homes and clinics.

Besides two dominant applications of mobile AR including hardware-based mobile AR and app-based mobile AR, another promising mobile AR applications is web AR. The main drawbacks of hardware-based mobile AR are its implementation cost and less flexibility. On the other hand, app-based AR applications are not suitable for a cross-platform deployment since they require extra downloading and installation in advance. In contrast, web AR is capable of providing pervasive mobile AR experience due to cross-platform service provisioning and the lightweight features of the web \cite{Qiao2019Proc}. However,
for the implementation of web AR in practical scenarios, several challenges exist in terms of energy efficiency, computational efficiency and networking. To enhance the performance of web AR applications, computational offloading to the cloud could be an advantageous option, but this may introduce additional communication delay, which may impact the user's experience. The spectrum efficiency and latency targets by the upcoming B5G networks as well as the emerging technologies including network slicing, D2D communications and multi-access edge computing, can be promising enablers for the deployment of web AR.








\section{Enabling Technologies for Wireless Tactile Internet}
\label{sec: sec6}

\subsection{Review of Main Challenges}
\label{sec: sec61}
In Table \ref{tab: TIResearchchallenges}, we have highlighted the key research challenges to be addressed for the practical realization of TI in future wireless networks across various application domains of TI. The main challenges concerning the connectivity issues include achieving ultra-low latency of $<1$ ms, ultra-high reliability of $>99.99\%$, high data-rate for some TI applications in the order of Gbps to Tbps and very high backhaul bandwidth. In addition to the above connectivity challenges (as described earlier in Section III),  the haptic domain presents several other challenges including: (1) the selection of suitable haptic codecs to capture the remote interactions, (2) the investigation of suitable coordination mechanisms and interfaces for H2M communications and suitable modalities for interactions, ensuring the stability of haptic control loop, (3) the investigation of suitable area-based or distributed sensing methods, (4) the investigation of methods for effective multiplexing of multi-model (haptic, audio and visual) sensory information, and (5) the design of novel performance metrics to characterize the system performance.

In the domain of Wireless Extended Reality (XR), i.e., wireless AR/VR, the additional challenges on top of the above mentioned connectivity challenges include balancing the quality-rate-latency trade-off, dealing with the scalability and heterogeneity, enhancing the localization and tracking efficiency, selecting between in-VR and in-network computation and investigating novel information theoretic principles to characterize the wireless XR systems. Furthermore, in the domain of autonomous, intelligent and cooperative mobility systems, as highlighted in Table \ref{tab: TIResearchchallenges}, the main challenges include ensuring low latency and high reliability of V2X services, dynamic route selection for the drones with low-latency, and efficient traffic management for Internet of drones. Also, suitable Machine Learning (ML)/Artificial Interference (AI) techniques for the prediction of movement/action to compensate for the physical limitations of remote latency, should be investigated. In the following, we provide some discussions on some of these challenges, along with the corresponding references.
\begin{table*}
\caption{\small{Research Challenges across various domains of Tactile Internet.}}
	\centering
\begin{tabular}{|l|l|}
\hline
TI Domain & Challenges \\ \hline
& Ultra-low latency ($<1$ ms) \\
& Ultra-high reliability ($>99.99$\%) \\
Connectivity  & High data-rates (Gbps-Tbps) \\
& Very high backhaul bandwidth \\ \hline
 &  Suitable haptic codecs to capture remote interactions  \\
& Coordination mechanisms/interfaces for human-to-machine interactions\\
Haptics & Stability for haptic control \\
& Area-based sensing and actuation \\
& Multiplexing of multi-model sensory information \\
& Novel performance metrics \\ \hline
& Quality-rate-latency trade-off \\
Wireless XR & Scalability and heterogeneity  \\
& Localization and tracking efficiency \\
& In-VR versus In-network computation \\
& Need for novel information theoretic principles \\ \hline
& Low latency and high reliability of V2X services \\
& Dynamic route selection for drones  \\
Autonomous, Intelligent and &  Traffic management for Internet of Drones \\
Cooperative Mobility Systems   & Design of ML/AI techniques for prediction of movement/action  \\
 & Multi-vehicle perception/control for safety and traffic efficiency \\
 & in cooperative automated driving \\
\hline
\end{tabular}
\vspace{-15 pt}
\label{tab: TIResearchchallenges}
\end{table*}

Providing stringent QoS for TI services in the upcoming 5G and beyond wireless networks is a crucial research challenge. 
To ensure the ultra-low round-trip latency and ultra-high reliability for each short packet in TI application scenarios, the main delay components including transmission delay, processing delay and queueing delay should be bounded with a small violation probability \cite{Cshe2016VTC}. These delays can be minimized at some level by using a short transmission frame structure, a short TTI and short codes including polar codes. Furthermore, the channel access delay at the access network and the latency involved in the backbone network could be minimized by utilizing novel access schemes and by adapting the network architecture, respectively.

Another issue is to provide the desired QoS to the TI services without significantly degrading the spectral efficiency and energy efficiency requirements of 5G and beyond networks \cite{Cshe2016energyefficient}. However, resource allocation needs to be conservative (for example, channel inversion power allocation with unbounded power) to meet the strict delay requirements, leading to the waste of useful energy and degradation in energy efficiency. Although there are some attempts in enhancing the energy efficiency under the delay bound violation probability and queueing delay constraints \cite{Liu2014energyefficient,Cshe2015resourceallocation} in different system settings, designing energy-efficient resource allocation techniques for TI applications under strict QoS constraints of TI services is challenging.

Furthermore,  most of the existing wireless systems are designed to support the Human-Type Communications (HTC) traffic and may not be efficient to handle a massive number of short data packets generated from IoT devices in TI applications. In contrast to the transmission of long packets, short data packet transmission differs mainly in the following two ways \cite{Durisi2016IEEEproc,Sharma2019mMTC}. First,  the assumption in the existing transmission techniques is that the size of the control information (metadata) is negligible with regard to the size of the information payload. Nevertheless, for the case of short packet transmission, the metadata size becomes no longer negligible and the above assumption does not hold, thus resulting in the need of highly efficient encoding schemes. Secondly, for the case of long packets, the thermal noise and channel distortions average out due to the law of large numbers and it is feasible to design channel codes which can enable the reconstruction of the information payload with a high probability, however, this averaging does not apply for the case of short packets and the classical law of large numbers does not hold for IoT/TI applications \cite{Durisi2016IEEEproc}, resulting in the need of new information theoretic principles.

Moreover, due to challenging requirements (mainly due to the strict requirement of $1$ ms round-trip latency), many existing congestion control mechanisms, higher layer and link layer based retransmission techniques for enhancing the communication reliability may not be applicable in TI applications. In order to achieve low latency required for TI applications, it is crucial to reduce the time-consuming signalling caused by different factors such as data transfer errors or loss detection and correction, and congestion control \cite{Holland2016ICT,Sharmacommletter2018}. Moreover, towards enabling the real-time human-machine interaction in TI applications, the following two issues are critical in addition to enhancing the speed of haptic data transmission \cite{Fettweis2014}: (i) the reaction time of the haptic system should fit within the time window defined by the reaction time of the involved human sense, and (ii) the time-lag of the feedback for different senses needs to be imperceptible for the scenarios where multiple senses (audio, visual, haptic) are being involved in an interaction.

Although recent advances in network architecture/topology, protocols and device hardware can reduce the end-to-end delays in a communication network to some extent, the achievable delay is limited by the finite speed of light since it provides an upper bound for the maximum range of tactile communications. For example, the maximum distance between the steering/control server from the tactile interaction point is limited to $150$ km as the light speed is $300$ km/ms \cite{Fettweis2014tactile}. To address this limitation, significant innovations are needed in the existing technologies and infrastructure. One approach to overcome this fundamental limitation caused by the finite speed of light could be to employ suitable ML techniques to incorporate some intelligence at the edge-network in a way that a similar action can be autonomously taken while the actual action is on its way from the network-side \cite{Aijaz2017tactile}.

The IEEE P1918.1 Working Group on the TI has set the following requirements for a tele-operation scenario \cite{Sachs2019adaptive}: (i) At least 99.9\% of network reliability for the haptic, audio and video channels, (ii) The latency for high dynamic environments should be between 1 and 10 ms for the haptic channel and between 10 and 20 ms for the audio and video channels, while for the medium or low dynamic environments, the latency requirement  ranges between 10 and 100  ms for the kinesthetics or 100 and 1000 ms for tactile devices and  30-40 ms for the video channels and 50-150 ms for the audio channels, respectively, and (iv) The required network bandwidth for the haptic control and feedback channel is on the order of  $100-500$ packets per second without employing any compression techniques, and $1000-4000$ packets per second with compression, with 2-48 bytes packet size depending on the number of DoFs of the teleoperator.  Also, depending on the features of the microphones, cameras, display devices and speakers, the audio and video channels between the operator and teleoperator require a bandwidth of about $5-512$ kb/s and $1-100$ Mb/s, respectively.





In addition, wireless VR technologies should be able to make an efficient transfer of VR video streaming by satisfying the following three UH requirements: (i) ultra-high data rate, (ii) ultra-high responsive speed, and (iii) ultra-high transfer reliability. The main difference between the scenarios having a single VR client and multiple VR clients is on the inter-user effects on the parallel transfer in different domains including spatial, temporal and frequency \cite{Huang2018MAC}. In this regard, authors in \cite{Huang2018MAC} presented the design and optimization of a multi-user MAC scheduling technique for multiple wireless VR scenarios, with the objective of maximizing the number of concurrent VR clients while guaranteeing three UH QoS requirements. The proposed scheme is based on the following functions: (i) video frame differentiation and delay-based weight calculation, (ii) spatial-frequency user selection based on aggregate delay-capacity utility, and (iii) link adaptation with the dynamic BLER target. Via numerical results, it has been demonstrated that the proposed technique increases the maximum number of concurrently served VR users by about $31.6$\%, which is higher than the rate produced by the conventional method relying on the fixed BLER target-based link adaptation and the maximum-sum-capacity-based scheduling.
\subsection{Enabling Technologies}
\label{sec: sec62}
In Table \ref{tab: EnablersPHYmACTI}, we present the key enabling techniques towards supporting TI applications in 5G and beyond networks along with the underlying sub-methods. In the following, we discuss these enabling technologies along with a review of the related references.

\begin{table*}
\caption{\small{Existing physical/MAC and Network level enabling technologies for TI.}}
\centering
\begin{tabular}{|l|l|l|l|}
\hline
\textbf{Protocol Layer} & \textbf{Enabling Technologies} &  \textbf{Sub-methods}& \textbf{References} \\
\hline
   &  & Cross-layer resource optimization   &  \cite{She2016crosslayer,Gholipoor2018}  \\
    & &  Energy-efficient resource allocation  & \cite{Cshe2016energyefficient} \\
   &  & Haptic resource allocation  & \cite{Aijaz2018IEEE,Aijaz2016towards}   \\
  & Dynamic resource allocation techniques         & Multi-cell resource allocation  & \cite{Hytonen2017multicell} \\
 & & Joint uplink-downlink resource allocation & \cite{Cshe2018joint} \\
 &  &Traffic aware resource allocation & \cite{Ryan2015enterprise,Wong2017predictive} \\
 & &   Massive MIMO & \cite{Tarneberg2017massive} \\
 & Advanced signal processing techniques &  Channel reciprocity & \cite{Lichannelreciprocity2018}  \\
 & &  Area spectral efficiency optimization &  \cite{Yilmaz2015analysis}  \\
 & &  Burstiness-aware Bandwidth Reservation  &  \cite{Hou2018burstiness} \\
 &  &    Diversity for transmission reliability                        &  \cite{Sheglobecom2016,Gringoli2018WiFi} \\
 Physical and MAC Layers & Transmission and Link adaptation schemes      &  Coding and modulation & \cite{Yoomodcoding} \\
&  & Asymmetric Transmit-Windowing & \cite{Taheri2016asymmetric}  \\
& & Multicarrier waveforms &  \cite{Schaich2014waveform} \\
&   & HCCA MAC protocol    & \cite{Feng2017hybrid}   \\
&  & Control channel design  &  \cite{Ashraf2015controlchannel} \\
&  Multiple access and scheduling techniques  & Non-orthogonal Multiple Access (NOMA) &  \cite{Sun2018short}  \\
&   & Pilot/overhead minimization  & \cite{Aziz2016IEEE} \\
&  Latency reduction and reliability enhancement &  Latency reduction methods  & \cite{Parvez2018survey,Pocovi2016impact} \\
& &  Reliability enhancement techniques &  \cite{Condoluci2017soft,Mountaser2018reliable,KimprocIEEE,Shariatmadari2017control} \\
\hline
&   Software-Defined Networking (SDN)  &   &  \cite{Petrov2018IEEE,Szabo2015TI,Petrov2018endtoend} \\
& Network virtualization &  & \cite{Shafigh2017dynamic,Holland2016ICT} \\
& Network coding &  & \cite{Cabrera2019software,Gabriel2018network,ParkmultiTCP2015} \\
Network Layer and Cloud-Level  & Collaborative edge-cloud processing &  & \cite{Ateya2017multilevel,Braun2017study,Chowdhury2017collaborative,Taleb2017multiaccess} \\
& Caching techniques  &  & \cite{Parvez2018survey,Xuenergyefficient2019,Liproc5GTI} \\
& Machine learning/AI &  & \cite{Sheng2018scalable,Ruan2018machine,Oteafy2019leveraging} \\
& Network architectural aspects & & \cite{Simsek20165Gtactile,AIjazproc2019TI,Arjunendend2018,Chowdhury2018context} \\
\hline
\end{tabular}
\vspace{-15 pt}
\label{tab: EnablersPHYmACTI}
\end{table*}

\subsubsection{Physical and MAC Layer Solutions}
In this subsection, we discuss various physical and MAC layer solutions towards the efficient management of resources, latency minimization and reliability enhancement for TI applications.\\
\textbf{1. Dynamic Resource Allocation Techniques}: Authors in \cite{Cshe2016energyefficient} studied the mechanisms to maximize the spectral efficiency and energy efficiency for TI applications under the constraint of QoS by taking the transmission delay and queueing delay into account. It was shown that the queueing delay violation probability derived from the effective bandwidth can be utilized for Poisson arrival process and for more burstier arrival processes than the Poisson process in the applications demanding ultra-low latency. Also, Queue State Information (QSI) and CSI dependent resource allocation policies were proposed to avoid energy wastage caused by the conservative design of loose upper bound of the queueing delay violation probability.

The incorporation of tactile data in a typical 5G and beyond wireless network makes the system implementation challenging due to different delay requirements of the TI data than those of the traditional data \cite{Simsel20165G}. Also, TI with the $1$ ms latency can highly revolutionize the mobility sector in various applications including the safety of pedestrians, platooning of the vehicles, and remote driving \cite{Fettweis2014tactile}. In the mobility related applications, the current achievable delay with the V2V and the V2I is still more than $10$ ms and this is not sufficient to support the autonomous vehicles traffic. Therefore, it is crucial to optimize the delay in TI applications including autonomous vehicles and remote driving.

Authors in \cite{She2016crosslayer} studied the problem of cross-layer optimization by considering the error probability, the transmission delay and the statistical queueing delay to ensure the QoS requirements of ultra-low end-to-end latency and ultra-high reliability in a TI system. By considering the dependency of queueing delay violation probability, packet dropping probability and packet error probability on the transmission resource and the policy, a joint optimization of the above three parameters was considered towards minimizing the transmit power of the BS. Via simulation results, it was shown that the optimized probabilities are in the same order of magnitude and all these factors need to be considered in the design of TI systems. Also, authors in \cite{Gholipoor2018} considered a resource allocation problem with the objective of maximizing the uplink sum-rate of the conventional data while fulfilling the delay requirements of the tactile data in an SCMA-based system model consisting of both the tactile and conventional data.

Furthermore, investigating the potential of LTE-A based cellular networks for TI applications may provide significant insights towards the deployment of TI applications in the existing cellular networks. In this regard, authors in \cite{Aijaz2016towards} investigated the radio resource allocation problem to support haptic communications in the LTE-A based cellular networks by taking the requirements for radio resource allocation in haptic communications along with the constraints of downlink and uplink multiple access techniques. To reduce the complexity of the optimization problem, an optimal power control policy was applied for the resource block allocation by first decomposing the problem and then transforming it to a binary integer programming.

Moreover, the authors in \cite{Hytonen2017multicell} presented three different coordinated multi-cell resource
allocation techniques including narrowband muting, a single frequency network and macro-diversity with soft combining for 5G URLLC scenarios in a typical indoor environment. Via system-level simulation results, it was shown that the inter-cell coordination is a promising approach to enhance the transmission reliability without causing longer delays as in the case of retransmissions.

In addition, authors in \cite{Cshe2018joint} investigated the joint uplink and downlink resource allocation problem to minimize the total bandwidth under the strict constraints of end-to-end delay and packet loss probability. A statistical multiplexing queueing mode was employed to minimize the desired packet rate for guaranteeing the queueing delay at the BSs' buffers. Furthermore, a joint optimization of downlink and uplink transmissions delays, queueing delay and bandwidth allocation, were performed to minimize the total bandwidth required to ensure the desired end-to-end delay and overall reliability. This joint optimization problem was decomposed in two steps with the first step considering the optimization of bandwidth with the given packet loss and delay components, and the second step dealing with the optimization of downlink and uplink transmission delays and queueing delay for the given end-to-end delay requirement. Via the presented analytical and numerical results, it was shown that there exists a tradeoff between the downlink and uplink bandwidth, and the optimization of delay components leads to the saving of almost half of the total bandwidth as compared to the scenario which does not consider the optimization of delays.

Although audio/visual information provides the feeling of being present in the remote environment, only the exchange of haptic information (force, motion, texture and vibration) can provide the real immersion into the remote scenarios. This remote operation along with the exchange of haptic information can enable the formation of human-in-the-loop networks. The emerging 5G-enabled TI is expected to transform the traditional human-in-the-loop networks to the human-in-the-loop mobile networks by supporting haptic communications over the 5G networks. In this regard, authors in \cite{Aijaz2018IEEE} identified the main architectural features of the human-in-the-loop mobile networks and the key requirements of haptic communications from the perspective of radio resource allocation. Also, in contrast to the traditional multimedia communications where uplink and downlink sessions are independent, uplink and downlink sessions in haptic communications are coupled due to the involved bidirectional haptic information exchange, and the fact that the degradation of QoS in one direction also impacts the QoS in the reverse direction. In this regard, authors in \cite{Aijaz2018IEEE} studied two distinct types of joint resource allocation problems related to symmetric and perceptual coding scenarios in haptic communications.

The TI aims to drastically revolutionize various aspects of our daily life via new applications and services. However, it is crucial to carefully evolve the underlying communication infrastructure to support the TI applications requiring ultra-high reliability, ultra-low latency and carrier-grade (ulra-high) reliability. To enable the real-time haptic communications in TI applications, it is important to keep the end-to-end latency (i.e., system response time) in the order of 1 ms \cite{Fettweis2014tactile}. The contributing factors for this end-to-end latency can be divided into three main categories \cite{Simsel20165G,Wong2017predictive}: (i) user interface ($\approx 300$ $\mu$s), (ii) radio interface ($\approx 200$ $\mu$s), and (iii) end-to-end latency between the wireless access point/BS and the steering/control server ($\approx 500$ $\mu$s).

While considering the aforementioned latency contributing factors, it is also important to investigate novel techniques which can minimize the end-to-end latency in the link between the BS/access point and the control/steering server.
In this regard, optical fiber technology seems to be a promising solution, but needs to consider the involved delays for switching, signal processing and protocol handling. Also, the distance between the edge TI BS/access point and control/steering server is limited to only a few kilometers \cite{Simsel20165G} and it is important to investigate novel Local Area Network (LAN) technologies which can support the reliable transfer of sensor/steering/control information with very low latency in addition to the conventional high-bandwidth applications/services. However, existing LAN technologies based on copper or/and multimode fiber are not sufficient to handle these heterogeneous types of traffic with different bandwidth demands and QoS requirements.

In the above context, the Passive Optical Local Area (POL) technology has emerged as a promising solution to address the heterogeneous requirements of future LANs since it can concurrently support multiple services via time-division multiplexing by utilizing the Gigabit or Ethernet passive optical network technology \cite{Ryan2015enterprise}. Towards supporting TI traffic, time division multiplexing-based POL technology can be extended to time and wavelength division multiplexing-based POL. In this regard, authors in \cite{Wong2017predictive} examined the effectiveness of the time and wavelength division multiplexing-based POL by employing a TI-enabled dynamic bandwidth and wavelength allocation algorithm towards concurrently supporting the TI traffic and the traditional healthcare traffic. The proposed algorithm estimates the average bandwidth of the TI and non-TI bandwidth at each integrated wireless access point or optical network terminal and uses prediction mechanisms at the central office in order to predict and allocated wavelengths and bandwidth to the accumulated traffic. Also, the number of active wavelengths is dynamically varied to avoid the network congestion while fulfilling the end-to-end latency constraints.

\begin{figure}
	\begin{center}
		\includegraphics[width=3.6 in]{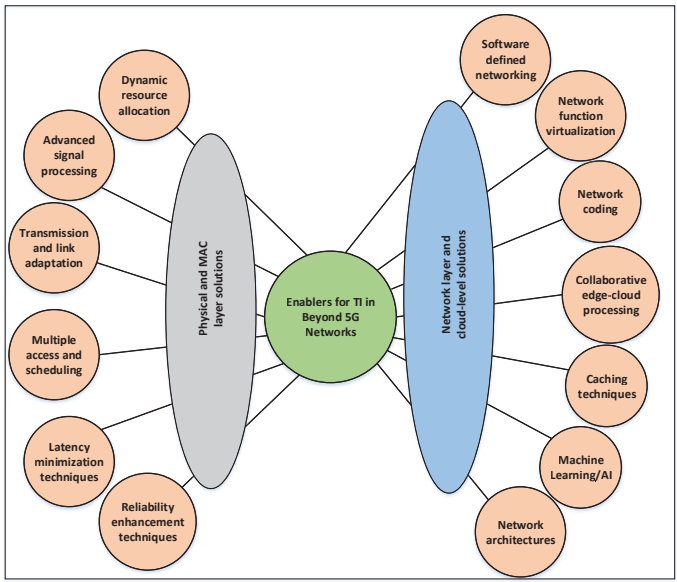}
		\caption{\footnotesize{Potential enabling techniques for Tactile Internet in 5G and beyond wireless networks.}}
	\label{fig: TIenablers}
	\end{center}
\end{figure}


\textbf{2. Advanced Signal Processing Techniques}:
The mission critical real-time applications over the TI such as tele-surgery demands for both ultra-reliable and low-latency requirements. The support for ultra-reliable and low-latency communications is considered as one of the crucial requirements of the upcoming 5G and beyond wireless networks. In this regard, authors in \cite{Tarneberg2017massive} investigated the utilization of massive MIMO technology for delivering URLLC in bilateral tele-operation systems as an integral component of the TI.


For the CSI acquisition in wireless communications systems, two types of training strategies are being utilized before data transmission: (i) downlink channel training strategy- in which the BS transmits pilot signals to the user, which then estimates the downlink CSI and feeds back to the BS, and (ii) uplink channel training strategy- in which the BS estimates the uplink CSI based on the pilot signals transmitted by the user and the corresponding downlink CSI is learnt based on the channel reciprocity. Although these techniques have been widely used in various conventional wireless systems, their performance degrades in wireless systems, which need to support short packets transmitted by the machine-type devices. In this regard, authors in \cite{Lichannelreciprocity2018} recently analyzed the performance of two training strategies by considering finite block-length constraints in short-packet transmissions, and derived closed-form expressions for the lower bounds on the achievable data-rates of these schemes. Furthermore, an approximate analytical expression for minimum channel reciprocity coefficient has been derived, and it has been shown that this coefficient decreases with the decrease in the block-length and with the increase in the number of transmit antennas. This result indicates that uplink channel training is more beneficial in short packet transmission with multiple transmit antennas.


Regarding the performance of short packet transmissions in Rayleigh block-fading channels, authors in \cite{Durisi2016shortpacket} investigated the performance tradeoffs among the key metrics including latency, throughput and reliability by considering a multiple antennas scenario. Also, authors derived the finite block-length bounds on the maximum
achievable coding rate over a MIMO Rayleigh block-fading channel without considering the prior knowledge of the CSI at the transmitter and receiver. The presented numerical results reveal that the conventional infinite block-length performance metrics including the ergodic capacity and outage do not provide an accurate estimate of the maximum coding rate
and they fail to characterize the tradeoff among throughput, latency, reliability and channel estimation overhead.

As a typical example of URLLC applications, factory automation can be considered, which demands for several new wireless technologies to enable discrete manufacturing, process automation and efficient production. It is very challenging to meet the demanding requirements of industrial automation in terms of ultra-high reliability and ultra-low latency \cite{Yilmaz2015analysis}. In order to analyze the performance of such industrial wireless networks, authors in  \cite{Chang2017QoS} defined a new performance metric (called QoS-constrained area spectral efficiency) by considering both the end-to-end delay and the reliability constraints into account. Subsequently, the authors solved an optimization problem to maximize the above metric with the power allocation by considering the constraints on the BS total power and QoS.

\textbf{3. Transmission and Link Adaptation Schemes}:
Ensuring URLLC between the remote device and the local operator is one of the crucial challenges in TI applications. The traffic of TI applications is predicted to be bursty \cite{Condoluci2017soft} and most existing works have not considered this burstiness aspect into account, thus resulting in either an underestimation of latency and packet loss probability in high traffic state or a sacrifice of the resource usage efficiency due to conservative resource reservation. To this end, authors in \cite{Hou2018burstiness} employed a Neyman-Pearson method to classify the packet arrival process into low and high traffic states, and then optimized the reserved bandwidth for these categories towards guaranteeing reliability and latency requirements. Via numerical results, the authors demonstrated that the proposed method being aware of the traffic states can save a huge amount of bandwidth as compared to the conventional case which is not aware of the traffic states. Also, the resulting bandwidth saving increases with the burstiness of traffic.

Furthermore, authors in \cite{Sheglobecom2016} considered the problem of designing the uplink transmission for TI scenarios, and analyzed the impact of spatial diversity and frequency diversity towards guaranteeing the transmission reliability. In this direction, a two-state transmission model was proposed based on the achievable rate with finite block-length channel codes. Also, by considering the system set-up, where multiple sub-channels were assigned to a device and each device selecting one sub-channel, various parameters such as the number of sub-channels, sub-channel bandwidth and the channel selection threshold, were optimized to ensure the transmission reliability.

In conventional sensor networks having the assumption of long packets and high duty cycles, various coding schemes such as
Bose-Chaudhuri–Hoccquenghem (BCH), low-density parity check, and convolutional codes can be utilized to enhance the energy efficiency. However, in many IoT sensory applications, the data packet size is very small and the duty cycle is also very low, therefore achieving higher energy efficiency becomes difficult due to various issues such as small coding gain, time synchronization and phase coherency. Due to low duty cycle, it becomes challenging to maintain time synchronization, and the required overhead can be prohibitively huge. Also, the resulted coding gain is smaller due to the short packet length. In this regard, authors in \cite{Yoomodcoding} studied the tradeoff between energy and bandwidth efficiency for short packet transmission in non-coherent transmission systems. Via numerical results, the authors showed that $16$-ary or $32$-ary orthogonal schemes are suitable candidate methods for short-packet transmission in wireless sensor networks. Also, the convolutional coding scheme was found to be a good candidate solution due to its simplicity, higher energy efficiency and robustness against practical imperfections including estimation errors.

With regard to low-latency and reliability requirements of 5G systems, there arises a fundamental tradeoff between latency and robustness of the system. For example, while employing the Out-Of-Band (OOB) emission techniques such as transmit windowing in OFDM-based systems, the OOB emission is suppressed, but on the other hand, the system becomes susceptible to channel-induced inter-symbol interference and inter-carrier interference. In this regard, authors in \cite{Taheri2016asymmetric} proposed to replace the widely-used symmetric window by an asymmetric window to improve the aforementioned latency-robustness tradeoff in LTE-based systems.

One of the promising methods to enable low end-to-end latency in TI applications is to employ short frame structures, but the employed finite block-length channel codes will result in transmission errors. This effect should be considered while satisfying the high reliability requirement. In this regard, authors in \cite{CShecrosslayer} studied cross-layer transmission optimization for TI applications by taking account of both the transmission delay and queueing delay in the end-to-end delay and different packet loss/error probabilities for the reliability. Mainly, a proactive packet dropping mechanism was proposed to satisfy the QoS requirement with the finite transmit power, and three different probabilities, namely, packet error probability, queueing delay violation probability, and packet dropping probability, were optimized to obtain the transmission and packet dropping policies for TI applications. Via numerical results, it has been illustrated that all the aforementioned probabilities are of the same order, and therefore, they all need to be considered equally while designing the transmission policy.

Traditional industrial automation systems are mainly based on wired communications, i.e., they use Fieldbus and Ethernet-based solutions. However, these wired solutions are expensive and cumbersome, and there is a need of replacing them with wireless alternatives. In the context of mission-critical communications for factory automation, providing ultra-reliable and low-latency communications is a crucial challenge. To this end, authors in  \cite{Jurdi2018variable} proposed a pilot-assisted variable rate URLLC method for the downlink of a factory automation network by considering a time-varying fading channel, and they demonstrated the significance of their proposed scheme (i) in extending the spectral efficiency range as compared to the schemes with the equal-rate transmission, and (ii) in providing robust ultra-reliability for a wide range of data payload sizes.

In addition to emerging 5G technologies, it is important to utilize the IEEE 802.11 (Wi-Fi) technology for achieving the low-latency communications in TI applications. In this regard, authors in \cite{Gringoli2018WiFi} studied the feasibility of using Wi-Fi in TI applications, and highlighted the potential of macro-diversity schemes for low-latency one-to-many communications via concurrent transmissions (C-Tx). More specifically, the authors analyzed the practical feasibility of using concurrent transmissions with IEEE 802.11 for both the DSSS and OFDM schemes. Also, the effect of limiting factors caused due to imperfect synchronization including concurrent carrier frequency offset, concurrent phase offset and concurrent timing offset has been investigated. The employed C-Tx flooding scheme in \cite{Gringoli2018WiFi} exploits the macro-diversity with the spatially distributed antenna arrays formed by combining the antennas of numerous devices to enable communications in the same time-frequency channel. This C-Tx flooding scheme provides several advantages including low-latency communications, reduction of overall network load/ congestion, lightweight operation, improved coverage and implicit time synchronization.

Another important aspect in short burst and low-latency communications is to design suitable physical layer waveforms. In this regard, authors in \cite{Schaich2014waveform} compared three candidate multi-carrier waveforms for the air-interface of 5G systems, namely, filtered Cyclic Prefix (CP)-OFDM, Filter Bank Multi-Carier (FBMC), and Universal Filtered Multi-Carrier (UFMC). The time-frequency efficiency of these techniques was analyzed by considering very small burst transmissions and with very tight latency constraints. Via numerical results, it has been shown in \cite{Schaich2014waveform} that the UFMC is the best choice in the short-packet and low-latency communications scenarios, especially in MTC and TI applications. Although the FBMC is efficient with long sequence transmissions, it suffers from the high time domain overheads and is subject to a significant performance degradation with short bursts/frames.

\textbf{4. Multiple Access and Scheduling Techniques}:
One of the important QoS criteria for the transfer of tactile information such as touch and actuation related information is latency, since they are very sensitive to the delay. In this regard, authors in \cite{Feng2017hybrid} considered a scenario having tactile body-worn devices connected via an IEEE 802.11 network, and analyzed the wireless communication latency involved while transmitting from these tactile devices to the access point by considering a Hybrid Coordination Function Controlled Channel Access (HCCA) MAC protocol. Furthermore, a theoretical expression for the average latency was derived by applying queueing theory.

In contrast to the existing LTE/LTE-A based wireless networks, future wireless networks need to be designed to support
emerging new use-cases which demand for ultra-high reliability and ultra-low latency. In this regard, authors in \cite{Ashraf2015controlchannel} investigated various design aspects of URLLC by considering factory automation as a use-case scenario. They showed that with the appropriate selection of coding, modulation, diversity technique and frequency/time resources, it is possible to have a fair balanced design of both the downlink and uplink control channels. Furthermore, by considering a similar factory automation example, authors in \cite{Ashraf2015controlchannel} investigated the feasibility of utilizing wireless communications for URLLC applications with some key design choices, and they showed the possibility of obtaining very low error rates and latencies over a wireless channel even in the presence of a fast fading channel, interference, antenna correlation and channel estimation errors. Also, it has been demonstrated via numerical results that promising enablers for high reliability and low latency are diversity and short transmission intervals without retransmissions, respectively.

Non-Orthogonal Multiple Access (NOMA) has been considered as one of the promising candidate techniques to address the scalability issue in handling the massive number of machine-type devices with short packets \cite{Sun2018short}. To this end, a trade-off among transmission delay, transmission rate and decoding error probability has been analyzed in two-users downlink NOMA system by considering finite block-length constraints. Another approach to enable short packet transmissions in future networks is to minimize the signalling overhead. To this end, authors in \cite{Aziz2016IEEE} proposed a signalling minimization framework in which a user can move freely within a dynamically allocated user-centric area and communicate with the network without the need of state transition signaling of the existing radio resource control. In addition, the packet transmission efficiency with short-packet transmissions can be enhanced by optimizing the pilot overhead. In this regard, authors in \cite{Mousaei2017pilot} studied the optimization problem of achievable rate as a function of pilot length, block length and error probability. The provided simulation results have revealed the significance of taking the packet size and error probability into account.

\begin{table*}
\caption{\small{Summary of latency reduction techniques in a wireless network.}}
\centering
\begin{tabular}{|l|l|l|l|}
\hline
\textbf{Network Segment} & \textbf{Main method} &  \textbf{Sub-methods} & \textbf{References} \\
\hline
    & Transmission frame design  & Short packets/Transmission Time Interval (TTI) & \cite{Durisi2016IEEEproc}  \\
       &   & Numerology and flexible sub-frame design & \cite{Lien20175GNR,Yazar2018flexibility} \\
       & Modulation and coding & Sparse Code Multiple Access (SCMA) & \cite{Moon2018SCMA} \\
       & & Finite block length coding & \cite{Huang2018efficient} \\
       &  & Polar coding & \cite{Zhang20165G} \\
RAN    & Transmitter design & New waveforms (UFMC, filtered CP-OFDM, FBMC) & \cite{Hammoodi2019green,Guan20175G} \\
       &         & Diversity (Interface, multiuser) & \cite{Nielsen2018URLLC,Sun2017exploiting}  \\
       &   & Asymmetric transmit-windowing    & \cite {Taheri2016GC} \\
       &  & Link adaptation and scheduling   &   \cite{Pocovi208joint}  \\
       & Control Signaling & Scaled control channel design & \cite{Ashraf2015control} \\
& & Sparse coding & \cite{JI2018ICC,Ji2019commlett}  \\
& & Outer-loop link adaptation & \cite{Ohseki2016fast} \\
& & Optimized access procedure & \cite{Jin2016access} \\ \hline
 & Optical wireless backhaul   & & \cite{Schulz2015low,Schulz2016robust} \\
 Backhaul & Edge caching & & \cite{Xuenergyefficient2019,Piao2019recent,Wang2016joint} \\
          & Coordinated and mobile-aware scheduling & & \cite{Andreoli2017mobile,Andreoli2017globecom} \\ \hline
  &  NFV-based methods & & \cite{Xiang2019reducing,Cziva2017edgeNFV}  \\
 Core Network       & SDN-based methods & & \cite{Xie2018cutting,Khalili2018flow} \\
       & Directory-based multicore architecture & &\cite{Asaduzzaman2017} \\
       & Fog/edge computing-based design & & \cite{Li2019service,Du2019enabling,Huang2017low} \\  \hline
\end{tabular}
\vspace{-15 pt}
\label{tab: latencyreduction}
\end{table*}

\textbf{5. Latency Minimization Techniques}: In wireless IoT systems, several factors across different protocol layers contribute to the overall end-to-end latency, thus leading to the need of promising physical layer and MAC layer techniques to support the low-latency transmission of TI applications over the wireless link. In the TI, some of the main delay components to be considered include transmission delay and queuing delay. If the average queuing delay is almost equal to the channel coherence time, the average transmit power may become unlimited. Therefore, the queuing delay is a key parameter of TI applications and should be considered in the TI implementation. This is especially the case when several types of data exist in the network, each of which has its own delay sensitivity.

In an LTE-based cellular system, the total latency can be divided into two major parts, i.e., User-plane (U-plane) latency and Control-plane (C-plane) latency \cite{Parvez2018survey}. The U-plane latency is measured by the directional transmit time of a packet to reach in the IP layer in between the User Equipment (UE)/Evolved UMTS Terrestrial RAN (E-UTRAN) node and the E-UTRAN node/UE. On the other hand, the C-plane latency corresponds to the transition time of a UE to switch from the idle state to the active state. Out of these two categories, the U-plane latency is considered important for latency minimization since the application performance is mainly dependent on this component. In addition, regarding the U-plane latency in a cellular network, it is contributed by various segments including RAN, backhaul, core network and Internet/data center.

Cross-layer interactions among different protocol layers are complex in wireless networks, and it is very important to address the issues associated with these cross-layer interactions from the perspective of sub-milliseconds latency.  Since end-to-end communications usually involves multiple wireless access technologies, as well as local area networks and core networks, achieving the desired latency for the transfer of audio, visual and haptic information is a crucial challenge \cite{Jiang2019lowlatency}. This end-to-end latency in wireless networks can be a function of various factors including MAC and networking technologies, link sharing, service/traffic demands and processing algorithms. Also, its contribution comes across different layers of the protocol stack including physical, data-link and networking, within a single network. If we focus on optimization of latency in a single layer, it may create harmful effects in other layers of the protocol. In this regard, authors in \cite{Jiang2019lowlatency} carried out the analysis and classification of various enabling technologies for achieving low-latency networking.

In Table \ref{tab: latencyreduction}, we list several latency reduction techniques which can be employed across different  segments of a wireless network, i.e., RAN, backhaul and core \cite{Parvez2018survey}, along with the related sub-methods and references. In the RAN segment, the main techniques include transmission frame design, efficient modulation and design, transmitter design and control signaling. Furthermore, the main latency minimization techniques concerning the backhaul segment include optical wireless backhaul, edge caching, coordinated and mobile-aware scheduling. Similarly, the main enabling techniques for latency reduction in the core segment include NFV-based methods, SDN-based methods, directory-based multicore architecture and fog/edge computing based design.

The total End-to-End (E2E) latency in a cellular network can be expressed as follows \cite{Parvez2018survey}:
\begin{equation}\label{}
T_{\mathrm{E2E}}=2 \times (T_{\mathrm{RAN}}+T_{\mathrm{Backhaul}}+T_{\mathrm{Core}}+T_{\mathrm{Transport}}),
\label{eq: E2E}
\end{equation}
where $T_{\mathrm{RAN}}$ is contributed by the UEs, eNodeBs and radio environment, and corresponds to the packet transmission time between the UEs and eNodeB. It occurs due to various physical layer operations including transmission time, processing time at the UE/eNodeB, retransmission time and propagation delay. Here, the $T_{\mathrm{RAN}}$ for a scheduled user can be expressed in terms of several contributors as follows \cite{Pocovi2016impact}:
\begin{equation}\label{}
T_{\mathrm{RAN}}=T_{\mathrm{Q}}+T_{\mathrm{FA}}+T_{\mathrm{tx}}+T_{\mathrm{bsp}}+T_{\mathrm{uep}},
\label{eq: E2ERAN}
\end{equation}
where $T_{\mathrm{Q}}$ denotes the queuing delay and depends on the number of
users that are multiplexed with the same radio resources, $T_{\mathrm{FA}}$ corresponds to the frame alignment delay and depends on the transmission frame structure and the involved duplexing modes, i.e., time division duplexing or frequency division duplexing. Similarly, $T_{\mathrm{tx}}$ denotes the time required for transmission processing and payload transmission while $T_{\mathrm{bsp}}$  and $T_{\mathrm{uep}}$ correspond to the processing delay at the eNodeB and UE, respectively.

In Equation (\ref{eq: E2E}), $T_{\mathrm{Backhaul}}$ corresponds to the time required to make connections between the core network, i.e., Evolved Packet Core (EPC) and the eNodeB, and this latency depends on the medium used, for example, microwave, copper or optical fiber. Similarly, $T_{\mathrm{Core}}$ refers to the processing time required at the EPC and this is mainly contributed by various network elements including SDN/NFV blocks, serving GPRS support node and Mobility Management Entity (MME). Furthermore, $T_{\mathrm{Transport}}$ corresponds to the data communication time between the EPC network and Cloud/Internet, and this is contributed by various parameters including the  distance between the server and the core network, the bandwidth and the employed communication protocol.

In order to reduce $T_{\mathrm{RAN}}$, several enhancements in terms of frame/packet structure, modulation and coding schemes, transmission techniques, waveform designs and symbol detection, need to be investigated. Furthermore, to reduce the $T_{\mathrm{Backhaul}}$, innovative methods including edge caching/computing and advanced backhaul techniques need to be investigated. Furthermore, SDN, NFV and ML-enabled intelligent techniques can be investigated to reduce the $T_{\mathrm{Core}}$, and the MEC-enabled cloud/caching/Internet can be utilized to minimize $T_{\mathrm{Transport}}$.

Furthermore, to minimize the latency in a wireless network, it is very important to make significant enhancements/modifications in the transmission strategy and frame/packet structure since there exists fundamental trade-offs among several parameters including latency, reliability, coverage, capacity and spectral efficiency. In the existing LTE-based system, latency is associated with the control overhead (about $0.3-0.4$ ms per transmitted packet) caused by various fields including transmission mode, pilot symbols and cyclic prefix \cite{Parvez2018survey}. Also, the retransmission time for each packet takes around $8$ ms, and the elimination of this period significantly affects the packet error rate.

\textbf{6. Reliability Enhancement Techniques}: Providing ultra-high reliability and ultra-low latency with the available limited bandwidth is one of the crucial requirements in TI applications. In the following, we discuss the existing works on various reliability enhancement techniques for URLLC scenarios including TI applications.

The existing multiple access control protocols in cellular networks are mostly based on the request and grant mechanism, i.e., the end-devices request for a scheduling request and then the BS provides a scheduling grant to the users. Although this Scheduling Request (SR) procedure can provide deterministic delays, it also introduces a significant delay due to the involved device-to-BS handshake procedure and may not be able to meet the requirements of TI applications \cite{Condoluci2017soft}. To address this, one of the potential solutions could be the reservation of some bandwidth for latency-sensitive users so that they can immediately transmit their data as soon as they have some data to transmit. Although such a bandwidth reservation method has been used in voice over Internet services having known and fixed transmit rate, the traffic pattern in TI applications is very bursty and irregular in nature \cite{Xu2017have} and this requires efficient bandwidth reservation techniques.

In the above context, authors in \cite{Condoluci2017soft} employed a soft resource reservation scheme for the uplink scheduling of LTE-based cellular users while addressing the particular requirements of haptic data traffic in cellular networks. In this soft reservation scheme, the uplink grant from one transmission is reserved for the subsequent transmissions in a softer manner without requiring the transmission of any new information from the device to the BS. Via numerical results, the performance of the proposed scheme was shown to be significantly better than the conventional SR scheme in terms of latency minimization. Furthermore, by considering the very bursty packet arrival processes in URLLC, authors in \cite{Hou2018burstiness} investigated the design of a spectrally-efficient resource management protocol by utilizing both data-driven and model-based unsupervised learning techniques with the objective of satisfying the required reliability and latency constraints while also minimizing the usage of bandwidth. Mainly, to address the issue of satisfying ultra-high reliability in the scenario with inaccurate traffic-state classification, an optimization problem was formulated with the objective of minimizing the reserved bandwidth under the latency and reliability constraints while also considering the classification errors. Via numerical results, it has been shown that the proposed burstiness-aware method can save about 40\%–70\% as compared to the traditional methods while also satisfying reliability and latency constraints.

One of the promising architectures for 5G and beyond networks is the Cloud-RAN architecture, which can offer flexibility in terms of moving baseband functionalities to the Central Unit (CU) which supports multiple remote Radio Units (RUs) or radio heads via Fronthaul (FH). In addition to other segments of the network, FH technologies should also be able to support the low latency and reliable communications for TI applications. The traditional FH consists of dedicated point-to-point links from the CU to each RU; but recently, a more economic solution of employing multi-path packet-based FH network instead of dedicated links has emerged  and this is more challenging in terms of meeting high reliability and low latency requirements of TI applications \cite{Mountaser2018reliable}. The latency requirements of FH links vary significantly between 55 $\mu$s and 1 ms \cite{Mountaser2017CRAN} depending on the split between the CU and the RU.

To fulfill these strict requirements, in contrast to the conventional solution of enhancing reliability with the feedback-based retransmission, transmission redundancy over multiple FH links between the CU and the RU seems to be an effective solution. In this regard, authors in \cite{Mountaser2018reliable} proposed to employ multi-path diversity over multiple FH links and the erasure coding of the MAC frames to achieve the high reliability and low latency transmission for TI applications in the C-RAN based wireless network. The authors investigated the average latency to achieve the reliable FH transmission as well as the reliability-latency tradeoff by using a probabilistic model. While considering the coexistence of enhanced Mobile Broadband (eMBB) and URLLC services, it was shown that efficient management of FH resources through orthogonal allocation can significantly lower the average latency and error probability as compared to the scenario where FH resources are allocated in a non-orthogonal mode.

The emergence of TI applications including haptic communications, immersive VR and cooperative automated driving has led to heterogeneous types of traffic having varying data rates and packet sizes, with different reliability and latency requirements. In this regard, it is crucial to develop novel technologies from the physical layer to the network layer in order to support these diverse traffics. In this regard, authors in  \cite{KimprocIEEE} discussed some promising physical layer technologies including multiple-access schemes, waveform multiplexing, synchronization, channel code design and full-duplex based transmission for spectrally efficient URLLC in TI services. Furthermore, to support the TI applications demanding high-reliability and low-latency, it is extremely important to categorize the traffic classes based on the traffic features and QoS requirements in a way that the traffic packets having the similar features and QoS can be incorporated in each sub-channel of a multiple access technique. To this end, authors in \cite{KimprocIEEE} have provided typical traffic characteristics and QoS for different TI applications including immersive VR, teleoperation, automotive, Internet of drones, and haptic interpersonal communications.

The reliability target of data transmission $\rho$ is related with the maximum BLER as: $\rho=1-\mathrm{BLER}$ and 3GPP targets to achieve the reliable communications of $10^{-5}$ BLER with $1$ ms radio latency in future cellular systems \cite{3GPPtechnical2017}. Towards enabling URLLC applications including industrial automation, e-Health, TI and vehicular communications in future B5G communication systems, it is essential to meet the reliability requirements of both the control and data channels. In this regard, authors in \cite{Shariatmadari2017control} developed general communication models  for both the uplink and downlink URLLC scenarios by taking the data and control channel errors into account. The proposed models considered the reliability constraints on  the control information which define the conditions under which a particular service can be supported. However, satisfying the reliability constraints under the low SNR is challenging and  several enhancement techniques in terms of channel estimation and Channel Quality Indicator (CQI) reporting, SR, Resource Grant (RG), and Acknowledgement (ACK)/Non-Acknowledgement (NACK) detection have been suggested in \cite{Shariatmadari2017control} to relax the strict constraints on the control channels. In Table \ref{tab: controlchannel}, we list the control channel reliability enhancement methods under the aforementioned steps of radio access and channel estimation procedure in cellular networks \cite{Shariatmadari2017control,Li20185G}.

\begin{table*}
\caption{\small{Techniques to enhance control channel reliability in URLLC applications including TI \cite{Shariatmadari2017control}.}}
\centering
\begin{tabular}{|l|l|l|}
\hline
\textbf{Protocol Step} & \textbf{Methods} &  \textbf{Principle} \\
\hline
                                                          & Power boosting for sounding reference signal  & Enables more accurate channel estimation in the uplink.   \\
Channel estimation and reporting    &  CQI backoff & Compensates the CQI decoding error at the eNodeB.   \\
                                                      & Adaptive configuration of CQI report    & Improves the reliability of CQI decoding. \\
                                                      & Outerloop link adaptation &  Enhances the accuracy of CQI reporting. \\ \hline
 Scheduling request   & Adaptive thresholding for energy detection & Detection threshold can be set based on channel conditions.  \\
            &  Inclusion of time-stamp in SR & Relaxes the reliability requirement for the SR signal \\
             & Repetition of scheduling information & Enhances the reliability of SR \\ \hline
 Resource grant   & Semi-persistent scheduling & Relaxes the reliability constraint for the downlink RG \\ \hline
 ACK/NACK detection & Asymmetric signal detection &  Helps to reduce the NACK missed detection. \\
 & Extra resource allocation for NACK & Improves the detection of NACK signal. \\ \hline
 \end{tabular}
\vspace{-15 pt}
\label{tab: controlchannel}
\end{table*}


Current LTE/LTE-A based cellular networks use feedback-based link adaptation techniques, i.e., Hybrid Automatic Repeat Request (HARQ) based on incremental redundancy in which the Modulation and Coding Scheme (MCS) to be used for transmission is first selected based on the present channel condition and in case the transmission fails, subsequent retransmission is performed by using a more conservative MCS method. Nevertheless, this feedback-based link adaptation approach may increase the communication delay due to the need of additional transmission time intervals while waiting for the ACK/NACK message. In addition to the increased latency, the spectral efficiency performance is also impacted due to continuous exchange of the CSI and large signalling/retransmission overheads per data packets while employing feedback-based link adaptation schemes in mMTC and URLLC systems. To this end, feedback-less of one-shot transmission schemes can avoid the signalling overheads involved in feedback-based link adaptation schemes; however, there appears a tradeoff between the reliability and the achieved latency. In this regard, the 3GPP proposes to utilize a scheduling-based radio access in the 5G New Radio (NR), and it is crucial to perform the radio access configuration at the 5G NodeB in order to dynamically switch between the feedback-based and feedback-less transmissions. To this end, authors in \cite{Lien2017efficient} analyzed the tradeoffs between feedback-less transmission and feedback-based transmission methods, and proposed an RL-based multi-armed bandit approach to harmonize the feedback-less and feedback-based transmissions by utilizing the underlying exploitation-exploration tradeoff.

Furthermore, another critical issue in B5G URLLC-based TI systems is to enhance the energy efficiency without making a sacrifice on the end-to-end latency. To this end, authors in \cite{Mukherjee2018energy} discussed several trade-offs between the user-plane delay and energy efficiency for future URLLC systems while considering both the network infrastructure energy efficiency and URLLC device energy efficiency. Regarding the infrastructure energy efficiency, the BS on-off switching, retransmissions and distributed network architectures have been discussed. Similarly, for the energy efficiency enhancement of URLLC devices, various techniques including discontinuous reception, power saving mode, mobility measurements and handover process, were discussed.

Moreover, authors in \cite{Li20185G} discussed various technologies for minimizing the latency and for enhancing the reliability in the URLLC systems. For the latency minimization purpose, various approaches including novel numerology and TTI duration, scheduling policies such as non-slot/mini-slot based scheduling and preemptive scheduling, and grant-free transmission, can be investigated. Similarly, for the reliability enhancement, signal power enhancement with micro-diversity techniques, HARQ enhancement and interference management techniques have been suggested. Furthermore, for enhancing the reliability of control channels, several methods including higher aggregation levels at the physical downlink control channel transmission, repetition of scheduling information, asymmetric detection of ACK/NACK and the adaptive configuration of CQI report, can be investigated.

Existing cloud-based computational solutions may suffer from the long communication delay as well as a high probability of error due to imperfections in the involved wireless links. To this end, Mobile Edge Computing (MEC), which brings the computing/processing units more closer to the user, has emerged as a  promising solution to reduce the delay caused by the existing cloud computing based solutions. However, different edge nodes may have varying channel qualities and the available radio frequency to be allocated for offloading to the MEC nodes is limited. Furthermore, it may be challenging to ensure a sufficient reliability concurrently while reducing the involved computational and communication latencies. To this end, authors in \cite{Liu2018ofloading} formulated a joint optimization problem to minimize the latency as well as the offloading failure probability and proposed three algorithms based on heuristic search, semi-definite relaxation and a linearization technique to solve the formulated optimization problem.

\textbf{7. Short-range Communication Technologies}: In contrast to the eMBB, the main performance metrics for URLLC in emerging 5G scenarios such as industrial automation  include low-latency and ultra-reliability. The conventional broadcasting strategy in an industrial factory may not be efficient to satisfy the URLLC requirements due to the massive number of actuators to be supported. In this direction, authors in \cite{Liu2018D2D} proposed a two-phase transmission protocol by forming clusters of D2D networks to enable URLLC in 5G wireless systems.

M2M is another promising short-range communication technology for providing ubiquitous connectivity among devices without the human control. The upcoming cellular technologies and infrastructures are expected to support M2M communications; however, several challenges in terms of supporting the massive number of connections, diverse QoS requirements of different applications, energy efficiency and spectral efficiency, need to be addressed \cite{Ghavimi2015M2M}.
Furthermore, congestion and system overloading are other crucial issues since the concurrent transmissions from a huge number of M2M devices may create congestion in both the random access and traffic channels \cite{CollabroativeRACH,Sharmacommletter2018}. For the provisioning of M2M communications over the 3GPP LTE/LTE-A based cellular networks, three main deployment paradigms have been defined by the 3GPP, namely, direct model, indirect model and hybrid model \cite{3GPParchitect}. In the direct model, the Application Server (AS) directly connects to an operator to communicate with the M2M devices without using the services of any external Service Capability Service (SCS). Besides, in the indirect model, the AS  is connected to the operator networks indirectly via the services of SCS to utilize the additional value added services of M2M. Finally, the hybrid model concurrently utilizes both the direct and indirect models for connecting the M2M devices to an operator network by using direct communications or an SCS.

\subsubsection{Network Layer and Cloud-Level Solutions}
In the following, we discuss various network layer and cloud-level solutions towards supporting TI applications in B5G wireless networks.\\
\textbf{1. Software Defined Networking}: The transfer of high-rate mission-critical traffic in the upcoming 5G and beyond networks needs the support from multiple radio access technologies and demands for dynamic orchestration of resources across both the core network and radio access segments. It is envisioned that 5G systems will be able to offer the features of network slicing, multi-connectivity and end-to-end quality provisioning mechanisms for transferring critical data within a software-controlled network. In this regard, the contribution in \cite{Petrov2018IEEE} introduced a softwarized 5G architecture to enhance the end-to-end reliability of the mission-critical traffic along with the corresponding mathematical framework.

Furthermore, authors in \cite{Szabo2015TI} considered the problem of reducing the communication latency in TI applications and demonstrated the significance of network coding in improving the communication latency and in reducing the packet re-transmission frequency. Also, a prototype to demonstrate the seamless integration of network coding with SDN towards a code centric network architecture was presented by implementing encoders, recoders and decoders as virtual network functions.

In order to guarantee the end-to-end reliability for the transfer of high-rate mission-critical traffic in 5G mobile networks, the network should be capable of supporting multiple radio access technologies and should perform dynamic allocation of resources across the core network and radio access segments. One of the promising platforms to achieve the desired end-to-end reliability in 5G networks is to utilize  the concept of network softwarization to form a softwarized 5G architecture. In this regard, authors in \cite{Petrov2018endtoend} employed SDN and Network Function Virtualization (NFV) technologies to form  an NFV-based SDN architecture in which all 5G network functions run as components of the software and the data plane is controlled by the SDN-like features.

\textbf{2. Network Virtualization}: Network virtualization is another promising technology enabler for supporting TI in B5G wireless networks. The most challenging requirements of TI applications are to meet the latency requirement of about less than $1$ ms and ``five-nines'' reliability, i.e., a one in a million chance of failure. One of the TI scenario examples which need such stringent requirements is the remote medical operations in which a surgeon needs to feel the sensation of touch and implement control actions by using force-feedback via the haptic clothing and other equipments. To this end, authors in \cite{Holland2016ICT} proposed the combined use of virtualized sub-GHz radios and licensed mobile radio to realize the TI in medical imagery applications. Also, the proposed concept was verified by employing Ofcom TV whitespace as the sub-GHz medium.

In addition, authors in \cite{Shafigh2017dynamic} proposed  a flexible cloud-based RAN for TI applications by utilizing the principle of dynamic network slicing, where the UE traffic can be temporarily offloaded from the operator's provided networks to the user's provided network whenever needed. Subsequently, a two-step matching game was employed for traffic-aware resource allocation in the proposed architecture with dynamic network slicing and load balancing; also offloading principles were used to support the concept of flexible RAN. Via numerical results, the authors demonstrated the significance of this deployment to the network operators and the end-users over the conventional CRANs.

\textbf{3. Network Coding}: Network coding can be considered as a promising technique to effectively manage the distributed communications and caching resources with a single code structure \cite{Cabrera2019software}. With regard to TI applications, network coding will facilitate the network operator in its task of increasing the reliability of its storage data against the failure of caching or computing nodes. The data reliability can be enhanced via some redundancy, which can be in the form of coded information or simple replication. Although the replication process requires the transfer of minimal traffic in case of a node failure, different storage nodes are needed to replicate the stored data. On the other hand, the conventional block codes including Reed–Solomon codes can drastically reduce the storage cost as compared to that of the storage required by the replication approach. However, all the information needs to be transferred over the network to recover the lost information, thus leading to a storage-repair traffic tradeoff. To this end, network coding can enable the operation of the system at the optimal points by balancing the storage-repair traffic tradeoff \cite{Firzek2016network}. The re-coding capability of network coding enhances the information reliability as well as balances the storage-repair tradeoff without the need of costly investments in the network architecture \cite{Cabrera2019software}.

Furthermore, authors in \cite{Gabriel2018network} utilized network coding to optimize the transmission reliability while taking the unknown and varying channel capacity into account. The  multipath TCP discussed in the literature utilizes multiple paths  to improve the throughput and reliability performance, but it suffers from the higher delay due to the employed Automatic Repeat Request (ARQ) as in the conventional TCP. To address these issues, the utilization of different multiple TCP schedulers \cite{Paasch2014} and rate allocation to minimize the latency \cite{ParkmultiTCP2015} have been discussed in the literature. In contrast, authors in \cite{Gabriel2018network} considered the network coding-based approach with the fixed-rate application traffic and time-varying channel capacity. The proposed method compensates the loss of capacity with the diversity of the paths and it is not dependent on the feedback required to maintain the reliability. An optimization framework was formulated to find the rate allocation for the paths which minimizes the error on all the configurations of channel capacities and the performance was evaluated with the help of real-world measurements.

\textbf{4. Collaborative Edge/Fog and Cloud Computing/Processing}: In contrast to the traditional Internet which focuses on transferring audio and visual senses, TI aims to transfer touch and actuation in the real-time. The MEC can be considered as one of the enablers for TI to reduce the round trip latency and perform the computational offloading from the cellular network. Furthermore, the MEC can provide the cloud-like functionalities including caching and computational resources at the edge-side of the network in order to reduce the latency and to better utilize the backhaul and core network resources \cite{Taleb2017multiaccess}. The MEC can benefit from the already existing ubiquitous coverage of cellular networks to support IoT and TI applications. It can facilitate multi-service and multi-tenancy operations to allow third-parties to utilize the edge-storage and  computing facilities by offering an open radio network edge platform. In \cite{Ateya2017multilevel}, the authors proposed a novel approach for a multi-level cloud-based cellular system in which each small-cell is equipped with edge computing capabilities via its connection with a micro-cloud unit. In the considered multi-level cloud architecture, micro-clouds are connected to the mini-clouds and ultimately to the main central-cloud. Due to the trade-off involved between the number of mini-clouds and their costs, it is necessary to optimize the number of mini-clouds and micro-clouds connected to the mini-cloud.

Various emerging future applications including industrial Internet, smart grids and driverless cars simultaneously demand for high bandwidth, resilience, low latency communication and security. Also, there is a growing need of shifting the store-and-forward networks towards the mobile edge cloud in order to enable the movement of the cloud as the device moves across the network. This novel concept of mobile edge cloud will enable advanced migration techniques and will support latency-sensitive services. In this regard, the authors in \cite{Braun2017study} proposed an agile cloud migration method for mobilizing the real-time and latency-sensitive applications frequently and quickly between multiple servers over the Internet.

In TI applications related to robotics and human-to-robot communications, limited energy, computing  power and storage resources become problematic, and efficient task allocation strategies need to be investigated. One of the promising approaches in this regard is task offloading to the collaborative nodes for enhancing the energy efficiency and task execution time. To this end, authors in \cite{Chowdhury2017collaborative} investigated a task allocation strategy by utilizing the conventional cloud, decentralized cloudlets and neighboring nodes to perform computational offloading. They have developed an analytical framework to compute the energy consumption, task response time and task allocation delay in integrated fiber-wireless multi-robot networks.


\textbf{5. Caching Techniques}: To enable the support of URLLC applications, it is crucial to redesign the wireless network architecture, including the RAN, the core network and the backhaul links between the RAN and the core network. Some of the promising enablers include SDN, NFV, edge-caching and MEC, and these technologies can contribute in reducing the latency by providing seamless operation and independence from the hardware functionality \cite{Parvez2018survey}. Furthermore, new physical air interfaces with small packet sizes,  small transmission interval time, new waveforms, MCSs, need to be investigated for latency minimization. Other emerging methods including radio resource optimization, carrier aggregation, massive MIMO and priority based data transmission need to be addressed towards this direction.

Moreover, proactive in-network caching has been receiving a significant attention in information-centric networking in order to reduce the latency of the upcoming wireless systems \cite{Parvez2018survey}. This caching method mainly involves four types of caching, namely, local caching, D2D caching, small BS caching and macro BS caching.  Out of these, local caching refers to caching at the user devices and D2D caching relies on D2D communications within the small cells. Similarly, small BS and macro BS caching refer to caching at the respective BSs. Furthermore, hybrid caching schemes can be exploited by combining the aforementioned caching schemes. In this regard, authors in \cite{Xuenergyefficient2019} devised an energy-efficient hybrid caching scheme  for TI applications along with a cache replacement policy to match this hybrid caching scheme by considering the Zipf distribution-based popularity of cached files. Via numerical results, it has been shown that the proposed hybrid scheme can reduce the latency while enhancing the overall energy efficiency for supporting TI applications in B5G networks. In addition, predictive caching can help to overcome the range limit of TI services due to the finite speed of light (the TI service range is limited to 100 km while using fiber) \cite{Liproc5GTI}.

\textbf{6. Machine Learning (ML)/Artificial Intelligence (AI) Techniques}: One of the main issues in TI is to understand the context of the remote terminals and the involved interactions, which is referred to as tactile cognizance  \cite{Oteafy2019leveraging}. Understanding such contextual information helps in delivering the perceived real-time operation in addition to the quality and depth of the haptic feedback. To this end, AI could be significantly useful in TI applications for different purposes including the following \cite{Oteafy2019leveraging}: (i) predicting the movement and action for compensating the physical limitations of remote latency, (ii) predicting the inference of movement and action for long-term skill acquisition and replication, and (iii) predicting reliable communication parameters.

In the above context, authors in \cite{Sheng2018scalable} proposed a scalable intelligence-enabled networking platform to remove the traffic redundancy in 5G audio-visual TI scenarios. The proposed platform incorporates a control plane, a user plane, an intelligent management plane and an intelligence-enabled plane. Out of these planes, the intelligence-enabled plane comprises a novel learning system that has decision making capability for the generalization and personalization in the presence of conflicting, imbalanced and partial data.
Furthermore, authors in \cite{Ruan2018machine} studied the application of ML intelligence in taking effective decisions for dynamically allocating the frequency resources in a heterogeneous fiber-wireless network. More specifically, they investigated the utilization of an artificial neural network for the following purposes: (i) in learning network uplink latency performance by utilizing diverse network features, and (ii) in taking flexible bandwidth allocation decisions towards reducing the uplink latency.

The upcoming 5G networks need to meet several requirements in terms of scalability, security,
low latency, mobility on demand and compatibility. However, the current IP address-based Internet architecture may not be capable of meeting these requirements due its inherent limitations in supporting on-demand mobility and scalable routing. In this context, authors in \cite{Sheng2018scalable} proposed a scalable intelligence-enabled networking, which is realized by utilizing information-centric networking schemes including identifier-locator mapping, routing and forwarding, to test the performance in typical 5G scenarios such as enhanced Mobile Broadband (eMBB), URLLC and mMTC. In addition,  an intelligence-enabled plane is also introduced by decoupling the functions of learning and the decision making from the management and control planes.

\textbf{7. Network Architectural Enhancements}: To address the challenging requirements of TI such as ultra-reliable and ultra-responsive network connectivity, many segments of the current mobile communication network infrastructures need to be significantly redesigned. Several enhancements need to be carried out including the design at the chip/circuit level, the redesign of the PHY/MAC layer solutions and cloud-computing solutions. From the perspective of backhaul/fronthaul designs, current individual and incompatible fronthaul/backhaul systems need to be migrated towards an integrated and flexible cross-haul network.

Furthermore, in contrast to the conventional Internet system which is designed to mainly transfer audio and visual information, TI should also be able to provide a medium to transport touch and actuation in the real-time. In other words, the architecture should be able to support both haptic and non-haptic information via the Internet, requiring the need of significant enhancements in the current network architecture. The end-to-end architecture of a basic TI system can be divided into three main domains, namely master domain, network domain and a slave domain \cite{Simsek20165Gtactile}, which are briefly described as follows.

As discussed earlier in Sec. \ref{sec:_sec24}, the master domain of the TI architecture is comprised of a human as an operator along with an HSI which converts the human input to the tactile input by using different tactile encoding techniques, and a machine controller \cite{AIjazproc2019TI}. A haptic device acts as an HSI and it enables a user to touch, feel and manipulate the objects in the virtual and real environments via the exchange of haptic information. In addition, the motion control devices can perform similar remote interaction functions without the exchange of haptic information, and the perceptual performance is enhanced via auditory and visual feedbacks \cite{AIjazproc2019TI}. Also, multiple human controllers can be involved in the master domain of some applications in order to carry out the collaborative operation of the slave domain. On the other hand, the machine controller comprises some digital controllers for industrial applications with several control algorithms with or without feedback.

Another component slave domain contains a tele-operator, i.e., a controlled robot, which interacts with several objects in the remote environment and is directly controlled by the master domain via various commands. In the context of NCSs, the slave domain comprises a system of sensors and actuators, in which the sensors measure the system state while the actuators adapt the system based on the commands obtained from the controller. On the other hand, the network domain provides a medium of bidirectional communication between master and slave domains, i.e., the interactions of a human to a remote environment. With the help of command and feedback signals, a global control loop is closed with the information exchanged between the master and controlled (slave) domains \cite{Simsek20165Gtactile}. TI applications need ultra-reliable and ultra-responsive connectivity to enable the efficient and timely delivery of sensing/actuation, control and touch information in the real time. Due to these stringent requirements of TI applications, current wireless networks should be substantially modified in terms of functional, architectural and protocol stack aspects.

One of the emerging concepts in the field of TI is tactile cyber-physical system, which has a wide variety of application scenarios including connected cars, smart grids, telemedicine, telepresence and interactive applications having actuators and sensors. In this regard, authors in \cite{Arjunendend2018} provided a high level block diagram of tactile cyber physical systems consisting of three main components, namely, an operator end, a teleoperator end and a tactile support network. They also discussed various underlying challenges from the system design perspective.

Moreover, one important design approach of TI  is the Human-Agent-Robot Teamwork (HART) \cite{Bradshaw2012human}. In contrast to the conventional approaches which divide the work only between machines and humans without creating any synergies, HART-based TI approach enables humans and robots to act as intelligent multi-agent systems, by exploiting the heterogeneous features of cognitive and physical tasks with the help of the cloud-center and dencentralized cloudlet resources. In such applications, collaborative robots (so-called cobots) can sense the surrounding environment  and cooperate with other robots in contrast to the conventional stand-alone robots, which are incapable of sense the environment. However, for the efficient design of an HART-centric approach, various contextual aspects related to the task (deadline, task size, task-type), the mobility, the latency, availability and capability of the collaborative agent/robot needs to be properly considered. In this regard, authors in \cite{Chowdhury2018context} studied the context-aware task migration techniques for efficiently performing real-time collaborations among the cloud/cloudlets, collaborative robots and human mobile users in the converged fiber-wireless communication infrastructures. In addition to the selection of a suitable cobot, the scheme suggested in \cite{Chowdhury2018context} deals with the migration of a task from one collaborative node to another towards reducing the task execution latency.

\section{Security and Privacy for TI Applications}
\label{sec:_sec7}
In the 5G-based wireless TI system, it is highly important to investigate techniques which can mitigate the radio jamming attacks since the wireless medium is intrinsically vulnerable to these attacks. The intelligent jamming attacks in wireless systems can be mainly categorized into two types, namely, control channel attacks and data channel attacks \cite{Liproc5GTI}. The jamming of uplink or/and downlink control signals can result in the failure of the transmissions and receptions of packets over the data channels.

The security threat in IoT systems including TI may arise in any phase of the data processing including data acquisition, information filtering, data fusion, representation, modeling, processing and interpretation \cite{SKSIEEE2017}. The main security issues in IoT systems include terminal security (e.g., illegal intrusion in device access, operation and control), data transmission security (e.g., stealing or loss of data due to unsafe transmission protocols), data processing security (e.g., revealing personal information) and management security (e.g., scalability of security techniques with massive data) \cite{Li2017blockchain}. As compared to the conventional scenarios, privacy loss may occur with a very high probability in wireless IoT systems due to various factors including higher level of interaction with the smart devices, location-based services and low awareness level on the user-side, thus leading to the investigation of novel privacy preserving mechanisms \cite{Porambage2016quest}. Also, due to the massive scale and distributed nature of IoT networks, maintaining the privacy of IoT data flowing in the network and fusion of data with privacy preservation are crucial aspects to be considered.

The conventional  security policies are usually centralized and are not effective in maintaining the security of distributed and heterogeneous IoT systems. In this regard, several security enhancement and privacy preservation mechanisms/frameworks have been recently suggested for IoT applications \cite{Shariatmadari2017control,Chow2017lastmile,Zhou2019effect,Sollins2019,Frustaci2018,Du2018distributed,Xiao2018IoTsecurity,Alrawais2017fog,Yin2018location,Cao2018privacy}. In Table \ref{tab: IoTsecurityreview}, we provide the key contributions of these works and their main focus on privacy or security or both.

\begin{table*}
\caption{\small{Review of recent contributions in the area of IoT security enhancement and privacy preservation.}}
\centering
\begin{tabular}{|l|l|l|}
\hline
\textbf{Reference} &  \textbf{Main focus}  & \textbf{Key contributions}\\\hline
\cite{Chow2017lastmile} & Privacy   & Discusses the privacy stack framework from IoT systems to the users \\ \hline
\cite{Zhou2019effect}  & Security and privacy & Discusses and analyzes the IoT security and privacy issues by considering 8 main IoT features  \\ \hline
\cite{Sollins2019}   & Security and privacy  &  Proposes a three-part decomposition process for IoT Big Data management  \\ \hline
\cite{Frustaci2018}   & Security   & Proposes a taxonomy analysis of the IoT security panoroma from perception, transportation and application perspectives \\ \hline
\cite{Du2018distributed} & Privacy  & Surveys IoT privacy preserving techniques from the perspective of data aggregation, trading, and analysis \\ \hline
\cite{Xiao2018IoTsecurity}   &  Security     & Identifies and reviews IoT attack models and learning-based IoT security techniques \\ \hline
\cite{Alrawais2017fog}  & Security and privacy   & Reviews the associated challenges and proposes fog computing based mechanisms \\ \hline
\cite{Yin2018location}   & Privacy       & Proposes a location privacy protection technique for data privacy in industrial IoT \\ \hline
\cite{Cao2018privacy} & Privacy   & Proposes a sparse coding based randomized response algorithm to achieve differential privacy \\ \hline
\end{tabular}
\vspace{-15 pt}
\label{tab: IoTsecurityreview}
\end{table*}

In the following, we discuss the key enabling technologies to enhance the security and privacy of TI applications.

\textbf{1. Blockchain}: A blockchain consists of a list of records or blocks which are linked to each-other by utilizing cryptography. In the emerging content-oriented networks, content providers may not have full control over their data due to high cost and complexity, and a large amount of information in the user's content, including  user name and location, can be learned easily if the information is not properly encrypted. To this end, it is crucial to develop effective access control and encryption techniques to preserve the underlying privacy information in future B5G networks supporting TI applications. The emerging blockchain technology can be considered a promising solution to achieve privacy preservation in future wireless systems since this distributed ledger system consisting of a number of blocks is considered to be public, secure and verified \cite{Zyskind2015privacy}. Each block of a blockchain contains a hash to be used for the integrity verification of the transaction information and the hash value of a block is dependent on the hash value of its preceding block.

Due to the openness and tamper-resistant features of the blockchain ledger, it can be useful to guarantee the privacy and access control of the content provider. To this end, authors in \cite{FanIETblcokchain} proposed a blockchain-based approach to address privacy issues in future content-oriented 5G networks by utilizing mutual trust between the users and the content providers. The blockchain has been combined with encrypted cloud storage based on the providers' data records present in the content-centric network. In this proposed approach \cite{FanIETblcokchain}, the tamper-resistant access policy on the public ledger enables the complete control of own information for each data owner. Furthermore, other advantages of this approach are that it enables reduced network congestion level, latency minimization and effective cost control. 

Although the blockchain technologies can enhance the degree of trust between devices in a decentralized network, some of its drawbacks include high equipment performance requirements and large-scale network formation. To this end, authors in
\cite{Li2017blockchain} suggested a multi-layer distributed model by combining the capabilities of the cloud-server and the reliability and security of IoT with the help of blockchain techniques. This blockchain-based multi-layer model is expected to provide several advantages, including the local coordination of IoT equipment with a centralized unit with improved safety, a reduction of the computational and network load of IoT with the help of multiple centers, and a secure and reliable IoT network with the contract records in multiple blockchains. On the contrary, such a model enhances the total network cost and also consumes much network resources and time to maintain the contracts \cite{Li2017blockchain}. In the similar context, authors in
\cite{Chen2018blockchain} proposed a joint cloud computing-based IoT service to address the privacy and data fusion issues of a single cloud approach used in most of the existing IoT services. This joint cloud computing based model in combination with blockchain can provide a trusted environment in order to exchange the resources among the clouds and to record all the involved transactions with the help of blockchain.

\textbf{2. Edge/Fog Computing}: Edge/fog computing is considered as a promising paradigm to extend the cloud functionalities (computing, data storage, communication, and networking) to the edge of the network; and it is significantly useful to support low-latency, mobility, location-awareness and geo-distributed applications \cite{Alnoman2019emerging,SKSIEEE2017}. However, the introduction of fog/edge computing brings the issues of security and privacy in IoT network including TI applications. Since the devices may malfunction or become susceptible to malicious attacks, it is highly essential to maintain a high level of trust between IoT devices and fog nodes so that the fog nodes can validate whether the service requesting devices are genuine or not \cite{Alrawais2017fog}. However, due to mobility issues and the lack of a centralized management, the existing trust models used in cloud computing platform cannot be directly applicable to fog computing scenarios. For example, the widely-used reputation-based trust models in e-commerce services may not be applicable for fog computing due to the dynamic nature of fog nodes and end-user devices in the fog layers. To this end, measuring the trust of a fog service and defining, verifying and monitoring the key attributes which characterize the trust of a fog service, are important aspects to be investigated.

Furthermore, it is important to ensure the security of the communications between fog nodes and resource-constrained IoT nodes, as well as communications among the distributed fog nodes. Also, since the fog nodes which are in the proximity to the end-users may collect sensitive information such as location, identity and utility usage, privacy preservation becomes challenging in fog nodes as compared to privacy preservation in cloud computing nodes. Moreover, since fog nodes may act as data aggregation and control point of the data gathered from the resource-constrained distributed IoT devices, authentication at different levels of gateways is also an important issue \cite{Mukherjee17security}.

\textbf{3. Machine Learning}: In IoT networks including TI, there may arise several security threats including Denial of Service (DoS) attacks, jamming, spoofing, man-in-the-middle attack, software attacks and privacy leakage. To minimize these threats and to improve the network security in wireless IoT networks, the emerging ML techniques seem promising. To this end, different ML techniques including supervised, unsupervised and reinforcement learning have been widely employed for access control, authentication, malware detection and anti-jamming offloading applications in different scenarios \cite{Xiao2018IoTsecurity}. However, the conventional ML techniques may not be directly applicable for the resource-constrained IoT devices and there exist several challenges in employing ML techniques in an IoT environment, including limited computational, radio and energy resources, distributed nature of devices, ultra-reliable and low-latency requirements for the learning outputs and the incorporation of heterogeneous traffic, including the traditional audio-visual, machine-type traffic and haptic traffic \cite{Sharma2019mMTC}. To this end, it is important to investigate innovative learning-based authentication, access control, secure offloading/storage for IoT/TI devices to protect from different attacks including physical and MAC layer, jamming, eavesdropping and smart attacks \cite{Xiao2018IoTsecurity}.

\textbf{4. SDN-based Design}: The SDN can greatly simplify the network management and control, and it can be a promising technology to enhance the security and privacy of IoT/TI applications. The integration of SDN in an IoT environment can provide the following advantages \cite{Sood2016SDN}: (i) The network's ability is enhanced by efficiently exploiting the under-utilized resources, (ii) Various steps of information processing in IoT networks including data/information acquisition, information analysis and decision making become simplified, (iii) The available network resources and access can be managed efficiently in a situation-aware manner to enhance the data link capacity between the users/devices, (iv) IoT networks can be made more scalable according to the demand and more agile with the help of an SDN platform, and (v) SDN-based traffic pattern analyzers will significantly help to simplify the data collection tools from the IoT devices. Furthermore, due to the decoupling of data plane from the control plane, hackers will have hard time accessing the data even if they become able to reach to the data plane. Also, SDN helps to automatically find an endpoint or a network segment impacted by the malware. More importantly, since SDN can have a global view of the underlying network including the mobility of nodes, traffic patterns and traffic volumes, it becomes simpler to implement security policies in SDN-enabled IoT networks than in the traditional networks \cite{Sood2016SDN}.

In the above context, some research works in the literature have studied the security aspects of SDN in IoT networks. Authors in \cite{Kartmakar2019} proposed a security architecture for IoT networks, which is capable of restricting the network access to unauthorized devices by utilizing the capabilities of SDN. Mainly, the integration of authorized flows in the IoT network and the authentication of IoT devices has been utilized to secure IoT networks from malicious devices and attacks. Furthermore, authors in \cite{Szymanski2017strengthening} proposed a combination of a centralized SDN control-plane, deterministic low-jitter scheduling, and lightweight encryption in layer 2 as a novel approach for enhancing the wireless security in emerging industrial IoT systems, including industrial IoT. With this approach, deterministic scheduling of the source, destination, time-slot and wireless channel can be achieved for every wireless transmission, and any unauthorized transmissions, even with a single packet, can be quickly detected in the layer 2 and this information can be sent to the SDN control plane for the corrective action.

\section{Open Issues and Future Recommendations}
\label{sec:_sec8}
In the following subsections, we discuss some open research issues and recommend some future research directions in various topics related to wireless TI.
\subsection{URLLC in TI Applications}
In URLLC applications of TI including immersive VR, teleoperation and cooperative automated driving, it is essential to effectively deliver haptic traffic (tactile and kinesthetic information) in addition to the conventional audio and video traffic \cite{KiMurllc}. Such heterogeneous and sporadic traffic types bring challenges in terms of latency and reliability requirements, thus leading to the need of novel technologies across the physical layer to the network layer. Furthermore, in contrast to the moderate reliability and very high rate for the visual perception, haptic perception needs fixed rate and high reliability. Therefore, in visual-haptic VR communications, there is a need to consider both  aforementioned objectives in a single underlying network. One promising technique to enable this is network slicing, which can create two different slices: one eMBB slice for the visual perception and a URLLC slice for the haptic perception \cite{Park2018URLLC}.

Balancing the tradeoff between energy efficiency and latency is another crucial aspect to be addressed in URLLC systems.
To this end, one interesting future research direction is to evaluate the energy efficiency and delay tradeoffs in the unlicensed spectrum (i.e., 60 GHz) since the continuous channel sensing required by the regulations in the unlicensed band causes a limit on the maximum transmission time \cite{Mukherjee2018energy}. Another important research question is how to enable the coexistence of TI systems operating under strict latency and reliability constraints with other systems such as eMBB systems demanding high throughput applications, and mMTC systems whose main target is to support a massive number of devices. How to efficiently multiplex URLLC traffic from TI systems and the traffic of other systems including URLLC and eMBB in both the downlink and uplink of future wireless systems is an important research issue to be addressed. Another crucial issue is how to prioritize the TI traffic with guarantees across both the access layer and network layer.

One promising approach suggested by the 3GPP for the multiplexing of URLLC and eMBB traffic in the downlink is preemptive scheduling \cite{3GPPNR2017}, in which the eMBB traffic is scheduled on all the available radio resources with a longer transmission time interval (i.e., 1 ms) and whenever the URLLC packet arrives at the 5G eNodeB,  it is transmitted to the respective user immediately by overwriting the ongoing eMBB transmission \cite{Li20185G}. However, the main issue with this approach is that the decoding performance of eMBB users gets degraded. To mitigate this issue, the 3GPP has introduced the parameter puncturing indication to inform the victim user which part of the transmission has been overwritten so that this effect can be considered during its transmission. Also, for the uplink multiplexing of the eMBB and URLLC traffic, the pause and resume method proposed by the 3GPP enables the URLLC user to take the already allocated resources from the eMBB user \cite{Li20185G}. In addition to inter-user multiplexing, intra-user multiplexing of eMBB and URLLC traffic could be another issue since the same user may use both the eMBB and URLLC services.

One of the promising future directions for enhancing the reliability of TI applications over wireless networks is to utilize the simultaneous connections with the help of multiple links. Also, another promising method is to utilize multiple paths for graph connectivity to avoid the single point of failure. Towards this direction, emerging graph signal processing techniques \cite{Ortega2018GSP} can be utilized to address various issues in large-scale wireless TI networks.
\subsection{Latency Reduction in Wireless TI Systems}
As highlighted earlier in Table \ref{tab: latencyreduction}, there exist several solutions towards reducing the latency in various segments of a wireless network. However, it is crucial to investigate novel methods across different segments of a wireless network to further reduce the latency  towards supporting TI applications in B5G networks. In the RAN segment, although millimeter wave (mmWave) communications seems to be a promising technology, several issues including non-line-of-sight beamforming, channel modeling with delay spread and angular spread, doppler, multi-path, atmospheric absorption and reflection/refraction/attenuation need further attention \cite{Rappmillimeter}. Furthermore, due to small-packet transmissions in IoT applications including TI, the conventional assumption of averaging out of the thermal noise and distortion for large packet sizes does not hold. Therefore, it is an important future research direction to investigate suitable transmission strategies and channel modeling for short packets across diverse frequency bands. Moreover, in most of the OFDM-based systems, synchronization and orthogonality cause a bottleneck in obtaining a lower latency and it is important to investigate suitable non-orthogonal and asynchronous multiple access schemes and waveforms which do not require much coordination and can provide robustness in the presence of channel fading and varying interference conditions \cite{Schaich2014waveform,Parvez2018survey}. In addition, the investigation of low complexity antenna structures, beam-steering active antenna array with beam steering capability, efficient symbol detection including compressed sensing and low-complexity sensing techniques, is important for low-latency communications. Other important future aspects include the treatment of latency sensitive packets either by providing priority-based timely access to the latency sensitive packets or by the reservation of resources. In addition, it is crucial to investigate suitable load-aware MAC protocols for TI applications to achieve the desired ultra-low latency with ultra-high reliability.

In the core network side, the main issues include the management and orchestration of heterogeneous resources including computing, network and storage \cite{Casellas2016IT} in the SDN/NFV-based core network while also minimizing the latency, and the need of intelligent, dynamic and adaptive techniques for the effective utilization of the existing heterogeneous backhauling networks. Moreover, further analysis and understanding of various tradeoffs including memory versus data-rate, storage versus link load and capacity versus latency in caching-enabled mobile edge computing, is needed. In addition, some important research directions in the context of caching-enabled wireless networks include the investigation of the impact of caching size, location and wireless link parameters on the end-to-end latency, the design of suitable protocols for intra-cache communications and caching redundancy with the latency constraint, and handover management in the scenarios having diverse caching capability with lower latency \cite{Parvez2018survey}. Furthermore, to overcome the issue of range limitation due to finite speed of light, emerging ML/AI techniques could be incorporated into the edge-side of TI networks in a way that necessary actions can be autonomously taken without the need of waiting for the action result to propagate from the network.
\subsection{Dynamic Resource Allocation in TI Systems}
The effective allocation of the involved resources in TI systems while satisfying the stringent latency and reliability constraints calls for the investigation of novel resource allocation techniques. In contrast to the conventional design approach of independent resource allocation in the uplink and downlink, joint uplink and downlink design is important for TI applications consisting of a closed-loop \cite{Aijaz2018IEEE}. However, resource allocation in TI systems is an unexplored domain that needs significant future attention. Various resources including communication resources (backhaul/fronthaul/access bandwidth, transmit power, antennas, device battery), computing resources (processor speed, memory size, computation rate/power) and caching resources (storage) \cite{SKSIEEE2017} are distributed throughout the B5G wireless networks supporting TI applications and they need to be managed optimally in order to utilize the available resources efficiently.

One promising direction to manage the distributed resources is to employ edge/fog computing as well as the cooperation of distributed edge computing nodes having different levels of energy and computing resources \cite{Alnoman2019emerging}. In such distributed edge computing based TI systems, several challenges including the energy efficiency and computational efficiency of edge nodes, energy-efficient task offloading and service response time of the end-users, need to be addressed \cite{Xiao2018distributed}. Furthermore, collaborative edge-cloud processing seems promising in utilizing the advantages of both the cloud-computing and edge-computing paradigms \cite{SKSIEEE2017}. Moreover, efficient management of admission control and user association is another promising approach to reduce the communication latency in wireless TI systems since the users can be associated with a suitable BS/access point which is able to serve them with the lowest latency.

In addition, towards enabling the priority-based resource allocation to TI applications and tactile-specific scheduling techniques in B5G networks, the SDN-based design and network virtualization seem promising. These paradigms will enable the flexible management of network/resource slices along with the required isolation and customization across different user devices and heterogeneous applications \cite{Aijaz2017tactile}. In this direction, future research works should focus on investigating suitable bandwidth-based and/or resource-based slicing techniques, isolation techniques to isolate resources across user devices or applications, and priority-based resource allocation and scheduling techniques for haptic users/applications.
\subsection{Haptic communications}
In contrast to the traditional communications, haptic communications demand for some specific requirements in terms of symmetric resource allocation in both the uplink and downlink, joint resource allocation in both the downlink and uplink, the consideration of bounded delay and guaranteeing the minimum rate throughout the haptic session. Furthermore, as detailed earlier in Section \ref{sec:_sec3}, there exist several challenges in employing haptic communications in future B5G networks. Some of these challenges include the efficient design of haptic devices and haptic codecs, maintaining the stability for control loop system, the design of effective multiplexing schemes to handle the cross-modal asynchrony due to the heterogeneous requirements of visual, haptic and auditory modalities. Other issues for the realization of haptic communications include: (i) achieving ultra-high reliability and ultra-responsive connectivity, (ii) prioritized and flexible resource allocation, (iii) incorporation of area-based or distributed sensing, (iv) overlay creation for multiuser haptic communications, and (v) the definition of novel performance metrics.

To address these issues, suitable physical/MAC layer and cloud/network level based solutions (detailed in Section \ref{sec: sec62}) should be investigated. Furthermore, due to time-varying delays and packet losses, the system instability could be problematic in practical haptic communications systems. One of the promising solutions to address this issue could be to investigate the applicability of ML techniques for enhancing the stability of the haptic control.
\subsection{Wireless Augmented/Virtual Reality}
One of the main issues in wireless AR/VR is to investigate suitable B5G network architecture which can deliver AR/VR capability, and to identify the corresponding interfaces to connect the AR/VR devices/terminals with the wireless network \cite{Westpal2017}. Also, another interesting research issue is how to utilize the advantages of the emerging edge-caching, multicasting and edge computing technologies for wireless AR/VR applications. Furthermore, other issues include identifying the rate adaptation mechanism for AR/VR, satisfying the reliability requirements of AR/VR at the network layer, characterizing the motion prediction, adapting the transport protocols based on the user channel information, characterizing QoS on the real-time by utilizing the SDN-based platform, and defining the interactions with the SDN controller.

In contrast to the conventional VR systems, which are mostly deployed over wired systems, wireless medium can provide much higher freedom of movement of immersion. However, the deployment of VR applications in a wireless environment faces a number of challenges, including how to characterize the QoS of the VR systems, how to manage both the tracking and QoS over highly dynamic wireless channels and how to design high data-rate and low-latency links to the VR services. Furthermore, other issues related to compression, data analytics and signal processing, need to be further investigated for wireless VR/AR systems.

Furthermore, with regard to mobile AR and web AR, there exist several challenges including the requirement of powerful computing versus limited computing capability of mobile devices, the requirement of real-time perception versus network delay, very high energy consumption versus limited battery capability of mobile/edge AR devices and the requirement of pervasive promotion versus diverse infrastructures (data formats, operating systems, computing and display platforms
\cite{Qiao2019Proc}. Also, related industries, academia and standardization bodies need to work together to resolve the compatibility issues arising from the diverse nature of devices, applications and infrastructures, in order to realize pervasive mobile and web AR.
\subsection{Hardware, Computational and Network Architectural Issues}
It would be ideal to have a separate network for haptic communications to meet the necessary requirements, however, due to the involved capital and operating expenditures, this is not a feasible solution \cite{Aijaz2017tactile}. Therefore, future B5G networks should be designed in a flexible way that they can simultaneously support various vertical applications including haptic, M2M, V2V, conventional video/data/voice while satisfying their diverse QoS requirements. One of the promising approaches in this regard is network slicing, by which different network slices can be allocated to different vertical sectors and can provide network on demand functionalities. To realize such a virtual network over a common physical network, the emerging SDN and NFV technologies seem promising. Also, self-organizing networking solutions can be utilized to bring the SDN's principle of separating control and data plane to the RAN-segment of future cellular systems \cite{Arsan2015SDN}. Furthermore, most existing TI related works assume that both the master and controlled domains are attached to the same wireless network. However, in practice, the network domain may comprise different wireless networks, and this demands for suitable SDN and virtualization-based network architectures to effectively coordinate among these networks.

In order to design effective tactile sensing systems, various aspects related to sensor hardware such as sensitivity and ability to measure several parameters, the arrangement of tactile sensors and the physical issues including spatial resolution and conformability, need to be addressed. Also, future advances in nano-technology and fabrication mechanisms are needed to effectively integrate various sensing units and signal processing modules \cite{Zou2017novel}. In addition to the development of tactile sensors, it is crucial to develop low-complex and hardware friendly techniques to process the raw information captured by these sensors.

In terms of computation in TI applications, the main challenges include \cite{Oteafy2019leveraging}: (i) online processing of haptic feedback for near real-time interactions, (ii) in-field processing to reduce ingress transmissions, and (iii) support of computationally intensive AI processing and training at the tactile edge. One of the promising approaches for enhancing the computational capacity per unit area could be highly adaptive energy-efficient computing box vision, which can contain over $10^9$ computing nodes within a $10$ cm$^3$ space \cite{Fettweis2012pathways}, and which relies on ultra-short range optical wireless communication between chips inside a cluster of computing nodes.
\subsection{Joint Design of Control and Communication Systems}
Most of TI systems need the fusion of control and communication systems to meet the desired requirements in terms of latency, reliability and system's stability. In this regard, it is necessary to design communication network protocols for achieving real-time performance and also to design suitable control algorithms to compensate the imperfections introduced by the communication medium \cite{AIjazproc2019TI}. These aspects have been independently handled by the communications and control theory research communities, but sufficient interactions between these two communities are currently missing. In this direction, it is crucial to have an integrated approach to optimize the network-aware control and control-aware network with sufficient interactions between the two research communities. Furthermore, maximizing the user's QoE while concurrently guaranteeing the stable global control loop in the presence of time-varying communication delay is one major challenge to be addressed in the networked teleoperation system with the haptic feedback. This will require the investigation of suitable dynamic control switching strategies for teleoperation systems for maximizing the QoE in the presence of time-varying communications delay \cite{Xu2017have}.

Furthermore, in vehicular platooning communications, it is essential to get the real-time information from the neighboring vehicles via inter-vehicle communications, and also to employ a suitable control law to achieve the objective of obtaining inter-vehicle spacing within the same platoon \cite{Liu2017joint}. To achieve this goal, one of the future directions could be to investigate a joint design of advanced control mechanisms and efficient IVC protocol for the safety and stability of cooperative driving, and also for the reliable delivery of the required information with low latency.
\subsection{Advanced Signal Processing and ML for TI}
The signals generated by the tactile sensors are usually noisy, complex with higher dimensions, and include much unnecessary information \cite{Kroemer2011learning}. Therefore, it is crucial to investigate advanced signal processing techniques to process such complex and multi-dimensional signals and low-complexity ML techniques to extract meaningful information from the raw data acquired by the tactile sensors.

To address the issues of high complexity (bad models), inefficient computation and slow convergence, AI/ML has been attracting a lot of attention in the wireless community and also it seems very promising for TI applications. However, various challenges due to fast time-varying channels, short stationary interval, distributed data and limitations on comptuation/energy, need to be addressed. In haptic communications over TI, it is challenging to have stable haptic control due to time-varying wireless channel. In this regard, one of the promising approaches is to utilize the recently emerging concept of edge intelligence at the edge of TI networks. With the help of edge caching and user-specific traffic management, congestion as well as the latency of haptic communications can be reduced \cite{Aijaz2017tactile}. Also, by employing various emerging ML/deep learning techniques \cite{Sharma2019mMTC}, suitable predictive and extrapolative/interpolative ML modules can be employed near the edge of the network. With the help of such ML-enabled TI support engines, the stability of haptic communications can be enhanced and also the fundamental limitation caused by the finite speed of light can be overcomed, thus enabling high communication range between two tactile ends \cite{Aijaz2017tactile}. Furthermore, hybrid models which may combine both the data-driven and model-driven paradigms should be exploited for future B5G TI applications.

\section{Conclusions}
\label{sec:_sec9}
Future Beyond 5G wireless networks are expected to support a broad variety of emerging TI applications having diverse QoS requirements. For this, several non-conventional challenges resulted due to the need of communicating haptic information in addition to the conventional audio and visual traffics over wireless media have to be addressed. The main technical requirements in this regard include ultra-low latency, ultra-high reliability, very high data-rate, energy efficiency, spectral efficiency and network throughput. To this end, starting with the vision of TI and recent advances, this survey paper has presented a generalized framework for wireless TI comprising the main technical requirements, key application areas, a TI architecture and the main enabling technologies, and has provided a comprehensive survey on various aspects of wireless TI. Three main paradigms of TI, namely, haptic communications, wireless AR/VR and autonomous, intelligent and cooperative mobility systems have been extensively discussed, along with the existing state-of-the-art review, the main challenges and potential enablers. Furthermore, key enablers for supporting TI applications/users in B5G systems have been identified and discussed along with the main sub-techniques. Moreover, considering security and privacy as important aspects of the TI deployment, existing related works have been reviewed along with the promising enablers. Finally, some important research issues and interesting recommendations for future research directions have been provided.

\section*{Acknowledgement}
This work was partially supported by the National Science and Engineering Research Council of Canada (NSERC), Reference No. RGPIN-2017-04423 and the Ryerson University Faculty of Science Deans' PostDoctoral Fellowship Fund, held by the 2nd author, and by the European Research Council (ERC) project AGNOSTIC (742648).



\end{document}